\documentclass{article}
\usepackage{graphicx}
\usepackage{epstopdf}
\usepackage{amsmath}
\usepackage{geometry}
\usepackage{amsfonts}
\usepackage{amssymb}
\usepackage{color}
\usepackage{lscape}
\newcommand{\doublespacing}{\let\CS=\@currsize\renewcommand{\baselinesstrech}
	{2.0}\tiny\CS}
\begin{document}
	\newcommand{\bd}{\begin{document}}
		\newcommand{\ed}{\end{document}}
	\newcommand{\bc}{\begin{center}}
		\newcommand{\ec}{\end{center}}
	\newcommand{\bfr}{\begin{flushright}}
		\newcommand{\efr}{\end{flushright}}
	\newcommand{\lt}{\left}
	\newcommand{\rt}{\right}
	\newcommand{\vs}{\vspace}
	\newcommand{\hs}{\hspace}
	\newcommand{\beq}{\begin{equation}}
		\newcommand{\eeq}{\end{equation}}
	\newcommand{\lb}{\linebreak}
	\newcommand{\pb}{\pagebreak}
	\newcommand{\mb}{\makebox}
	\newcommand{\fb}{\framebox}
	\newcommand{\mc}{\multicolumn}
	\newcommand{\ben}{\begin{enumerate}}
		\newcommand{\een}{\end{enumerate}}
	\newcommand{\bit}{\begin{itemize}}
		\newcommand{\eit}{\end{itemize}}
	\newcommand{\oln}{\overline}
	\newcommand{\un}{\underline}
	\newcommand{\lefq}{\lefteqn}
	\newcommand{\ba}{\begin{array}}
		\newcommand{\ea}{\end{array}}
	\newcommand{\beqa}{\begin{eqnarray}}
		\newcommand{\eeqa}{\end{eqnarray}}
	\newcommand{\beqas}{\begin{eqnarray*}}
		\newcommand{\eeqas}{\end{eqnarray*}}
	\newcommand{\bfg}{\begin{figure}}
		\newcommand{\efg}{\end{figure}}
	\newcommand{\bds}{\begin{displaymath}}
		\newcommand{\eds}{\end{displaymath}}
	\newcommand{\btb}{\begin{tabbing}}
		\newcommand{\etb}{\end{tabbing}}
	\newcommand{\para}{\parallel}
	\newcommand{\pad}{\partial}
	\newcommand{\nn}{\nonumber}
	\newcommand{\la}{\leftarrow}
	\newcommand{\ra}{\rightarrow}
	\newcommand{\lgla}{\longleftarrow}
	\newcommand{\lgra}{\longrightarrow}
	\newcommand{\La}{\Leftarrow}\newcommand{\Ra}{\Rightarrow}
	\newcommand{\Lra}{\Leftrightarrow}
	\newcommand{\Lgla}{\Longleftarrow}
	\newcommand{\Lgra}{\Longrightarrow}
	\newcommand{\lan}{\langle}
	\newcommand{\ran}{\rangle}
	\renewcommand{\a}{\alpha}
	\renewcommand{\b}{\beta}
	\newcommand{\g}{\gamma}
	\newcommand{\G}{\Gamma}
	\renewcommand{\d}{\delta}
	\newcommand{\eps}{\epsilon}
	\newcommand{\Th}{\Theta}
	\newcommand{\s}{\sigma}
	\newcommand{\lam}{\lambda}
	\newcommand{\D}{\Delta}
	\newcommand{\ds}{\displaystyle}
	\newcommand{\vare}{E}
	\newcommand{\pr}{\prime}
	\newcommand{\ro}{\rho}
	\newcommand{\nab}{\nabla}
	\newcommand{\m}{\mu}
	\newcommand{\n}{\nu}
	\newcommand{\Sg}{\Sigma}
	\newcommand{\p}{\pi}
	\newcommand{\R}{I\!\!R}
	\newcommand{\om}{\omega}
	\newcommand{\Om}{\Omega}
	\newcommand{\ovra}{\overrightarrow}
	\newcommand{\ze}{\zeta}
	\newcommand{\vart}{\vartheta}
	\newcommand{\tri}{\triangle}
	\newcommand{\f}{\frac}
	\newcommand{\iny}{\infty}
	\newcommand{\pro}{\propto}
	\renewcommand{\arraystretch}{1.25}
		
	
	\bc {\huge Information theoretic measures within Schr\"odinger-Dunkl framework in spherical coordinates} \\	\vs{1cm} 	{\it Akash Halder and Amlan K. Roy{\footnote{Email: akroy@iiserkol.ac.in, akroy6k@gmail.com}}\\Department of Chemical Sciences, Indian Institute of Science Education and Research (IISER) Kolkata, Mohanpur-741246, Nadia, WB, India.\\Debraj Nath$^*${\footnote{Email: debrajn@gmail.com ($^*$Corresponding author)}} \\Department of Mathematics, Vivekananda College, Kolkata - 700063, WB, India.} \ec
		\bc {\large {\un{Abstract}}} \ec
	In this article, we have presented analytical solution of Schr\"odinger-Dunkl equation with Deng-Fan molecular potential in spherical coordinates. The ro-vibrational energy of some selected diatomic molecules ScH, TiH, VH and CrH are obtained under a simple, new approximation to the centrifugal term in presence of three reflection operators and Dunkl parameters. The angular wave functions are obtained in terms of Jacobi polynomial, whereas radial wave functions in terms of hypergeometric function. The analytical results of Shannon entropy, expectation, Heisenberg uncertainty, entropic moment, disequilibrium, R\'enyi entropy and Tsallis entropy of marginal density (radial $r$, and angular $\theta,\, \phi$) and total density functions are obtained in Schr\"odinger-Dunkl system with respect to the weighted Lebesgue measure. This has been possible by making use of factorization method for Shannon entropy. The absolute per cent deviation between the analytical and numerical results for all the information theoretic measures remain well within 0.0001\%. The effect of reflection operators on angular wave solutions and on information theoretic measures have been investigated. In essence, a number of statistical measures have been reported for Deng-Fan potential in the Dunkl-Schr\"odinger framework. \\
		\date{\today}


\section{Introduction}
In 1950, a novel quantization method was proposed by Wigner \cite{wigner1950}. A kind of deformation of Heisenberg algebra was introduced, where the product of Wigner parameter $\mu$ and parity operator in the traditional position-momentum communication relation, i.e., $[\widehat{x}, \widehat{p}]=i\hbar(1+ 2\mu \widehat{R})$, with $\widehat{R} f(x) = f(-x)$, does not change the equation of motion. This is done though the Dunkl operator \cite{dunkl1989, dunkl1991, dunkl2008}, which is expressed in terms of the differential-difference and reflection operators. The deformation of quantum mechanics has been a subject of considerable research in last few decades. These operators have emerged in a number of important areas of mathematical physics, such as in the analysis of polynomials in multiple variables having discrete symmetry \cite{dunkl2008, dunkl2014}, representation of complete set of quantum integrals in terms of Dunkl derivative \cite{heckman1991}, application of quantum Calogero-Sutherland model \cite{lapointe1996} in conformal field theory, bosonization of sypersymmetric quantum mechanics \cite{plyushchay1996}, anyons in (2+1) and (1+1) dimensions \cite{plyushchay1996a}, para-fields and para-statistics \cite{govorkov1983}. 

In literature, the most studied system is harmonic oscillator, due to its important connection to various quantum systems. The eigenfunctions of 1D Dunkl oscillator has been given in terms of $\nu$-deformed Hermite polynomial \cite{chung2019}. Solutions for (1+1)-dimensional Dirac-Dunkl oscillator has been found via the $SU(1,1)$ algebraic approach \cite{ojeda2020}. \emph{Exact} solutions of the \emph{generalized} Dunkl oscillator was presented \cite{dong2021}, along with the thermodynamic properties (Helmholtz free energy, internal energy, specific heat capacity, entropy). The thermal properties of relativistic Dunkl-Klein-Gordon and Dunkl-Dirac oscillators have been reported recently \cite{hamil2022}. The coherent states for real and complex variables by means of the Segal-Bargmann transformation of Dunkl-type has been published \cite{ghazouani2022}. The momentum representation with Dunkl derivatives in 1D has been presented by constructing Dunkl-Heisenberg relation \cite{chung2023}, through the $\nu$-deformed exponential and $\nu$-deformed trigonometric function. Recently various entropic measures such as R\'enyi entropy, Fisher information, disequilibrium, relative entropies, as well as associated Jensen divergences have been provided \cite{nath2024}. 

The superintegrability of isotropic Dunkl oscillator model in a plane has been investigated in Cartesian and polar coordinates \cite{vinet2013}. The coherent states and its time evolution in a plane was considered in a 2D Dunkl oscillator \cite{salazar2017}. The Dunkl-Klein-Gordon equation in 2D has been solved by an algebraic method \cite{mota2021}. The Landau levels of (2+1)-dimensional Klein-Gordon oscillator in presence of an external magnetic field were also pursued \cite{mota2021a}. Eigenfunctions and eigenvalues of 2D \emph{generalized} Dunkl-Klein-Gordon relativistic equation has been reported \cite{hassanabadi2022} via a Nikiforov-Uvarov (NU) method. Exact solutions are also reported for 2D Dirac Dunkl oscillator in presence of a uniform magnetic field \cite{mota2019}. 

Likewise, the isotropic 3D Dunkl oscillator is also well studied. In Cartesian, polar (cylindrical) and spherical coordinates, the relevant Schr\"odinger equation is \emph{exactly} solvable, admits separation of variables and maximally superintegrable \cite{genest2014}. The eigenfunctions are expressed in terms of generalized Hermite polynomials; the radial and angular solutions are obtained in terms of generalized Hermite, Laguerre and Jacobi polynomials. In parallel, the relativistic Dirac-Klein-Gordon oscillator in Cartesian and spherical coordinates become separable---the eigenfunctions are written in terms of associated Laguerre and Jacobi polynomials \cite{hamil2022a}. \emph{Exact} solutions of 3D generalized Dunkl oscillator in Cartesian coordinates are available \cite{dong2023}. 

Apart from harmonic oscillator, only very few works are available on other potentials of interest within the Dunkl framework. Only recently, some model pedagogic systems, such as particle in a 1D box, as well as radial and angular solutions of certain central-field problems, such as free particle spherical wave, pseudoharmonic oscillator and Mie-type potentials \cite{mota2022} have been considered. The Dunkl-Coulomb problem in a plane is maximally superintegrable and \emph{exactly} solvable  (as products of Laguerre polynomial and Dunkl harmonics on circle) \cite{genest2015}. The 2D Dunkl-Coulomb problem in polar coordinates has been \emph{exactly} solved and Perelomov coherent states are constructed for radial part \cite{salazar2018}. The energy spectrum as well as wave function of the 3D problem are obtained by means of a spectrum generating algebra technique based on so(1,2) Lie algebra \cite{ghazouani2019}. \emph{Exact} solution of the 2D Dunkl-Klein-Gordon equation has been offered for Coulomb potential \cite{mota2021}. Relativistic Klein-Gordon equation was solved for generalized Dunkl anharmonic oscillator in 2D \cite{hassanabadi2022}. Eigenvalues and eigenfunctions of Dirac-Klein-Gordon equation for Coulomb potential and Dirac fine structure has been published recently \cite{hamil2022a}. Using the Lewis-Riesenfeld method, exact solution of time dependent oscillator was defined within Dunkl formalism \cite{dunkl-oscillator-td}. 

From the above consideration, it appears that there is a lack of work on the Dunkl-Schr\"odinger equation for problems other than the few model potentials mentioned. The purpose of this communication is to offer such a treatment for the important Deng-Fan molecular potential, introduced nearly 70 years ago \cite{deng1957}. This three-parameter model potential has found considerable relevance in the parlance of molecular vibration. A characteristic feature is that in the limiting values of internuclear distance, this provides a correct qualitative asymptotic value. Attempts have been made to calculate eigenvalues and eigenvalues through a number of attractive elegant methods such as exact solvability approach \cite{mesa1998}, algebraic method \cite{codriansky1999}, supersymmetric shape invariance \cite{zhang2011}, NU method \cite{kdsen2013jmc, nath2021}, factorization method \cite{mustafa2015}, generalized pseudospectral method \cite{roy2015}, Feynman path integral formalism \cite{boukabcha2018}, exact quantization rule \cite{oluwadare2018}, etc. 

The Dunkl angular momentum operator contains the reflection operators and therefore, reflection operators have significant effecftive role on angular wave solutions \cite{genest2014,ghazouani2019,mota2022,dunkl-oscillator-td}. So, the reflection operators have important contribution on information theoretic measure of angular wave functions.

At first, the eigenvalues and eigenfunctions of the relevant Dunkl-Schr\"odinger equation of the Deng-Fan potential are obtained by means of the NU method, utilizing a simple yet effective modified Pekeris approximation for centrifugal term, recently proposed and succesfully applied by us. This is elaborated in Sec.~\ref{sec2.method}. This will extend the scope of application of the Dunkl operator in physical problems. Then a detailed analysis of energy with respect to Dunkl parameters is given for four diatomic molecules (ScH, TiH, VH, CrH), in ground and low-lying excited states (including both $\ell=0$ and $\ell \neq 0$ states). Results are presented in tabular and graphical format. Various expectations like $\langle r^n \rangle$ (for $n =-1, -2, 1, 2$) as well as Heisenberg uncertainty product is offered in Sec.~\ref{sec3.uncertainty}. Next, using this wave function, expression for radial, angular as well as composite Shannon entropy ($S$) (in position space) is derived with the help of weighted Lebesgue measure in Sec.~\ref{sec4.shannon}. The radial and angular quantities are found to be expressible in terms of hypergeometric function and Jacobi polynomial respectively. Next, entropic moment, disequilibrium, R\'enyi entropy and Tsallsi entropy are derived in Sec.~\ref{sec6.moment}. The effect of reflection operators on information theoretic measures are investigated. In all cases representative results are given for same four molecules. Finally a few concluding remarks are made in Sec.~\ref{sec7.con}.  \\

\section{Solution of Schrödinger-Dunkl equation in spherical coordinates}\label{sec2.method}
The time-independent Schr\"odinger-Dunkl equation in three dimensional cartesian coordinates can be written as, 
\beq
	\left[-\frac{\hbar^2}{2\mu}\nabla_D^2+V(\mathbf{r})\right]\psi(\mathbf{r})=E\psi(\mathbf{r}), 
\eeq 
where the Dunkl-Laplacian operator is defined by, 
\beq
	\nabla_D^2=D_x^2+D_y^2+D_z^2.
\eeq
The Dunkl derivatives along the three coordinate axes are given by, 
\beq
	\ba{ll}
	D_x=\frac{\partial}{\partial x}+\frac{\mu_x}{x}(1-\widehat{R}_x),
	D_y=\frac{\partial}{\partial y}+\frac{\mu_x}{y}(1-\widehat{R}_y),
	D_z=\frac{\partial}{\partial z}+\frac{\mu_x}{z}(1-\widehat{R}_z), 
	\ea 
\eeq 
where $\widehat{R}$'s denote the reflection operators along $x,y,z$ axes, as follows, 
\beq
	\ba{ll}
	\widehat{R}_xf(x,y,z)=f(-x,y,z),
	\widehat{R}_yf(x,y,z)=f(x,-y,z),
	\widehat{R}_zf(x,y,z)=f(x,y,-z),
	\ea
\eeq
with $\mu_x,\mu_y,\mu_z$ signifying the Dunkl parameters. Now using the following relations,  
\beq
	\ba{ll}
	x=r\cos\phi\sin\theta,y=r\sin\phi\sin\theta,z=r\cos\theta , 
	\ea
\eeq
the Dunkl-Laplacian operator, $\nabla_D^2$ can be written as \cite{genest2014,mota2022}, 
\beq
	\ba{ll}
	\nabla_D^2&=M_r-\frac{L_D^2}{\hbar^2r^2},
	\ea 
\eeq 
where $L_D^2=-\hbar^2\left(N_{\theta}+\frac{1}{\sin^2\theta}B_{\phi}\right)$ represents the angular Dunkl momentum operator, while $N_{\theta}$ and $B_{\phi}$ are given by the following expressions, 
\beq
	\ba{ll}
	M_r&=\ds\frac{1}{r^{2a}}\frac{\partial}{\partial r}\left(r^{2a}\frac{\partial}{\partial r}\right),~a=1+\mu_x+\mu_y+\mu_z,\\
	N_{\theta}&=\ds\frac{1}{\sin\theta}\frac{\partial}{\partial\theta}\left(\sin\theta\frac{\partial}{\partial\theta}\right)+2\left[(\mu_x+\mu_y)\cot\theta-\mu_z\tan\theta\right]\frac{\partial}{\partial\theta}-\frac{\mu_z}{\cos^2\theta}(1-\widehat{R}_z),\\
	B_{\phi}&=\ds\frac{\partial^2}{\partial\phi^2}-2\left(\mu_x\tan\phi-\mu_y\cot\phi\right)\frac{\partial}{\partial\phi}-\frac{\mu_x}{\cos^2\phi}(1-\widehat{R}_x)-\frac{\mu_y}{\sin^2\phi}(1-\widehat{R}_y).
	\ea 
\eeq 
Here the reflection operators, in the spherical coordinates, act as given below, 
\beq
	\ba{ll}
	\widehat{R}_x\psi(r,\theta,\phi)=\psi(r,\theta,\pi-\phi)=s_1\psi(r,\theta,\phi),\\
	\widehat{R}_y\psi(r,\theta,\phi)=\psi(r,\theta,-\phi)=s_2\psi(r,\theta,\phi),\\
	\widehat{R}_z\psi(r,\theta,\phi)=\psi(r,\pi-\theta,\phi)=s_3\psi(r,\theta,\phi).
	\ea
\eeq 
Now, let, 
\beq
	\psi(r,\theta,\phi)=\frac{R(r)}{r^a}H(\theta)G(\phi)
\eeq
be the solution of our desired Schr\"odinger equation,  
\beq
	H\psi(r,\theta,\phi)=E\psi(r,\theta,\phi).
\eeq 
Then one obtains, 
\beq\label{bphi}
	B_{\phi}G(\phi)=\eps^{(\phi)}_{m}G(\phi),
\eeq
\beq\label{ntheta}
	\left(N_{\theta}+\frac{\eps^{(\phi)}_m}{\sin^2\theta}\right)H(\theta)=\eps^{(\theta)}_{\ell,m}H(\theta),
\eeq 
and
\beq
	\ds\frac{d^2U}{dr^2}+\frac{2a}{r}\frac{dU}{dr}+\left[\frac{2\mu}{\hbar^2}(E-V)+\frac{\eps^{(\theta)}_{\ell,m}}{r^2}\right]U(r)=0,~U=\frac{R(r)}{r^a}\label{rse}.
\eeq 
In order to solve the $\phi$ equation, Eq.~(\ref{bphi}), we consider a transformation, $p=\cos2\phi$, leading to, 
\beq
	(1-p^2)\frac{d^2G}{dp^2}+\left[\mu_x-\mu_y-(\mu_x+\mu_y+1)p\right]\frac{dG}{dp}-\left[\frac{\mu_x(1-\widehat{R}_x)}{2(1+p)}+\frac{\mu_y(1-\widehat{R}_y)}{2(1-p)}+\frac{\eps^{(\phi)}_m}{4}\right]G=0.
\eeq 
Now let us substitute, 
\beq
	G(\phi)=(1-p)^{l_1}(1+p)^{l_2}F^{(\phi)}(p),
\eeq 
in the above equation. This gives the following equation, 
\beq
	(1-p^2)\frac{d^2F^{(\phi)}}{dp^2}+\left[\b_1-\a_1-(\a_1+\b_1+2)p\right]\frac{dF^{(\phi)}}{dp}+m(m+\a_1+\b_1+1)F^{(\phi)}=0,
\eeq 
where
\beq
	\ba{l}
	\a_1=2l_1+\mu_y-\frac{1}{2},\b_1=2l_2+\mu_x-\frac{1}{2},\\
	4l_1\left(l_1+\mu_x-\frac{1}{2}\right)-\mu_x(1-s_2)=0\implies l_1=\frac{1-s_2}{4},\\
	4l_2\left(l_2+\mu_y-\frac{1}{2}\right)-\mu_y(1-s_1)=0\implies l_2=\frac{1-s_1}{4}.
	\ea
\eeq 
After some algebraic manipulation, the final solution of the $\phi$ problem, Eq.~(\ref{bphi}), can be written as, 
\beq
	G_m^{(s_1,s_2)}(\phi)=N_{m}^{(\phi)}(1-p)^{l_1}(1+p)^{l_2}\,P_{m}^{(\a_1,\b_1)}(p),~p=\cos2\phi,
\eeq
and the corresponding angular energy is given by,  
\beq
	\eps^{(\phi)}_{m}=-4(l_1+l_2)(l_1+l_2+\mu_x+\mu_y)-4m(m+\a_1+\b_1+1),
\eeq
where $P_{m}^{(\a_1,\b_1)}(x)$ denotes the classical Jacobi polynomial. The normalization constant \cite{stegun}, 
\beq\label{nor.phi}
	N_m^{(\phi)}=\left[\frac{ (2m+\a_1+\b_1+1) m! \G(m+\a_1+\b_1+1)}{2^{2l_1+2l_2} \G\left(m+\a_1+1\right)\G\left(m+\b_1+1\right)}\right]^{\frac{1}{2}}
\eeq 
is obtained from the orthogonality condition, 
\beq
	\ds\int_0^{2\pi}G_{m}^{(s_1,s_2)}(\phi)G_{m'}^{(s'_1,s'_2)}(\phi)d\chi_{\phi}=\delta_{m,m'}\delta_{s_1,s_1'}\delta_{s_2,s_2'},~d\chi_{\phi}=|\cos\phi|^{2\mu_x}|\sin\phi|^{2\mu_y}d\phi.
\eeq
If $m\rightarrow m-l_1-l_2$, then $\eps_m^{(\phi)}\rightarrow-4m(m+\mu_x+\mu_y)$. In this case, $m\in \frac{1}{2}+\mathbb{N}$, when $s_1s_2=-1$, and $m\in \mathbb{N}\cup\{0\}$ when $s_1s_2=1$. But if $m=0$, then $s_1$ and $s_2$ must be equal to one \cite{genest2014}.\\ 
Similarly, to solve the $\theta$ problem, Eq.~(\ref{ntheta}), we let $q=\cos2\theta$, and $H=(1-q)^{l_3}(1+q)^{l_4}F^{(\theta)}(q)$, to yield, 
\beq
	(1-q^2)\frac{d^2F^{(\theta)}}{dq^2}+\left[\b_2-\a_2-(\a_2+\b_2+2)q\right]\frac{dF^{(\theta)}}{dq}+\ell(\ell+\a_2+\b_2+1)F^{(\theta)}=0,
\eeq
where
\beq
	\ba{l}
	\a_2=2l_3+\mu_x+\mu_y,~
	\b_2=2l_4+\mu_z-\frac{1}{2},\\
	4l_3\left(l_3+\mu_x+\mu_y\right)+\eps^{(\phi)}_m=0\implies l_3=\frac{1}{2}\left[-\mu_x-\mu_y+\sqrt{(\mu_x+\mu_y)^2-\eps^{(\phi)}_m}\right],\\
	4l_4\left(l_4+\mu_z-\frac{1}{2}\right)-\mu_z(1-s_3)=0\implies l_4=\frac{1-s_3}{4}.
	\ea
\eeq 
After some simplification, the solutions of Eq.~(\ref{ntheta}) can be written as, 
\beq
	H_{\ell,m}^{(s_3)}(\theta)=N_{\ell,m}^{(\theta)}(1-q)^{l_3}(1+q)^{l_4}\,P_{\ell}^{(\a_2,\b_2)}(q),~q=\cos2\theta,
\eeq
and the eigenvalues are obtained as, 
\beq
	\eps^{(\theta)}_{\ell,m}=-4\ell(\ell+\a_2+\b_2+1)-4(l_3+l_4)\left(l_3+l_4+\mu_x+\mu_y+\mu_z+\frac{1}{2}\right),
\eeq
whereas, the normalization constant is gotten as \cite{stegun}, 	
\beq\label{nor.theta}
	N_{\ell,m}^{(\theta)}=\left[\frac{\left(2\ell+\a_2+\b_2+1\right) \ell! \G(\ell+\a_2+\b_2+1)}{2^{\a_2+\b_2+1}\G(\ell+\a_2+1)\G(\ell+\b_2+1)}\right]^{\frac{1}{2}}
\eeq 
from the following condition, 
\beq
	\ds\int_0^{\pi}H_{\ell,m}^{(s_3)}(\theta)H_{\ell',m'}^{(s_3')}(\theta)d\chi_{\theta}=\delta_{\ell,\ell'}\delta_{s_3,s_3'},~d\chi_{\theta}=|\sin\theta|^{2\mu_x+2\mu_y}|\cos\theta|^{2\mu_z}\sin\theta\,d\theta.
\eeq
If $\ell\rightarrow \ell-l_4$ and $m\rightarrow m-l_1-l_2$, then $\eps_{(\ell,m)}^{(\theta)}\rightarrow -4(\ell+m)(\ell+m+\mu_x+\mu_y+\mu_z+\frac{1}{2})$ and $l_3\rightarrow m$. In this case, $\ell\in\mathbb{N}\cup\{0\}$ when $s_3=1$, and $\ell\in\frac{1}{2}+\mathbb{N}$ when $s_3=-1$ \cite{genest2014}.\\
Thus the energy of angular momentum operator, $L_D^2$ is $-\hbar^2\eps_{(\ell,m)}^{(\theta)}$. Then the radial Schr\"odinger equation becomes,  
\beq
	\ds\frac{d^2R}{dr^2}+\left[\frac{2\mu}{\hbar^2}(E-V)-\frac{\g}{r^2}\right]R(r)=0.
\eeq

\begin{figure}[t]
	\centering
	\includegraphics[width=15cm,height=10cm]{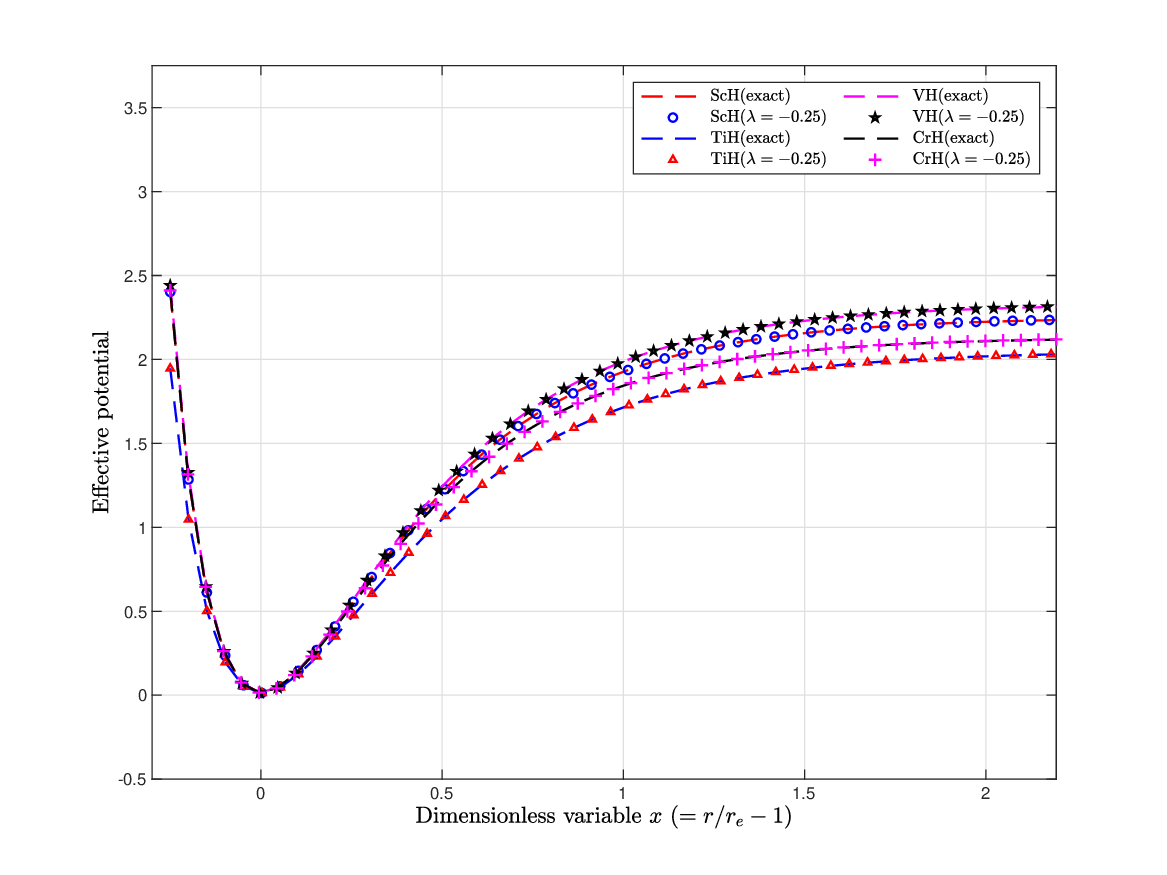}
	\caption{\label{fig1} Plot of effective potential, Eq.~(30), in presence of Dunkl-parameter. Dashed and marker lines represent this, with exact and approximate centrifugal terms. These are given for ScH, TiH, VH, CrH molecules.}
\end{figure}

In this work, we have considered the Deng-Fan molecular potential \cite{deng1957,kdsen2013jmc,mustafa2015,nath2021,boukabcha2018}, having the form, 
\begin{equation} \label{eq:2}
		V(r)=D_e \left( 1- \frac{b} {e^{\a r}-1} \right)^2, \ \ \ \ b=e^{\a r_e} -1,  
\end{equation}
where the three positive parameters $D_e, r_e, \a$ represent dissociation energy, equilibrium internuclear distance and radius of the potential well respectively. The \emph{effective} potential in presence of the Dunkl parameter is then given by, 
\beq
	V_{eff}=V(r)+\frac{\hbar^2\g}{2\mu r^2}, ~\g=a(a-1)-\eps^{(\theta)}_{\ell,m},
\eeq 
where $\frac{\hbar^2\g}{2\mu r^2}$ is the centrifugal term. 
To find the solution of radial Schr\"odinger equation we use a new approximation to the centrifugal term, recently proposed by us and found to be quite successful in a number of molecular potentials \cite{nath2021,nath2021a,nath2021b,nath2023}, 
\beq\label{comvex}
	\f{1}{r^2}\approx\a^2\left(d_0(\lam)+\f{d_1(\lambda)\,s}{1-s}+\f{d_2(\lambda)\,s^2}{(1-s)^2}\right),~s=e^{-\a r}
\eeq 
where
\beq
	d_0(\lam)=\f{\lam}{12}+\f{(1-\lam)c_0}{u^2},~d_1(\lam)=\lam+\f{(1-\lam)c_1}{u^2},~d_2(\lam)=\lam+\f{(1-\lam)c_2}{u^2}.
\eeq 
The coefficients $c_0,c_1,c_2$ are obtained by the Pekeris-type approximation \cite{badawi72, pekeris, mustafa2015,boukabcha2018}, where the potential is minimum at the point $r=r_e$. Thus one obtains, 
\beq
	\ba{ll}
	c_0&=\f{1}{u^2}\left[3-3u+u^2+\left(2u-6\right)s_e+\left(u+3\right)s_e^2\right],\\
	c_1&=\f{2}{s_eu^2}\left(1-s_e\right)^2\left((3+u)s_e+2u-3\right), \\
	c_2&=-\f{1}{s_e^2u^2}\left(1-s_e\right)^3\left((3+u)s_e+u-3\right),\\
	s_e&=e^{-u},~u=\a r_e.
	\ea
\eeq
Note that $\lambda=1$ resembles a Greene-type approximation \cite{greene} having well-behaved-ness near origin, whereas $\lambda=0$ corresponds to the Pekeris-type approximation (well behaved near $r=r_e$). For $0<\lambda<1$, it remains well behaved in domain $(0,r_e)$.

\begin{table}[t]
	\centering
	\begin{tabular}{rrrrr}\hline
			Molecule& \multicolumn{1}{c}{$D_e$} & \multicolumn{1}{c}{$r_e$}& \multicolumn{1}{c}{$\alpha$}& \multicolumn{1}{c}{$\mu$}\\
			&(eV)&(\AA) & $(\AA^{-1})$&(a.m.u)\\\hline
			ScH&2.25& 1.776&1.41113&0.986040\\
			TiH&2.05&1.781&1.32408&0.987371\\
			VH&2.33&1.719&1.44370&0.988005\\
			CrH&2.13&1.694&1.52179&0.988976\\\hline
	\end{tabular}
	\caption{\label{table1.parameters} Set of parameters of four diatomic molecules \cite{kdsen2013jmc}.}
\end{table}

Note that $\gamma$ depends on reflection operators and Dunkl parameters, and therefore, these parameters have significant effect on the centrifugal potential. In absence of reflection operators and Dunkl parameters, centrifugal term depends on the angular quantum number $\ell$. In our previous study we have applied a new approximation to the centrifugal term \cite{nath2021,nath2021a,nath2021b,nath2023}. In Fig.~\ref{fig1}, we have plotted the effective potential for ScH, TiH, VH and CrH diatomic molecules in presence of Dunkl parameters: $\mu_x=0.03,\mu_y=0.01, \mu_z=0.03$, $s_1=s_2=s_3=1$ under the new approximation, with $\lambda=-0.25$. The molecular parameters, adopted from \cite{kdsen2013jmc}, are listed in Table~\ref{table1.parameters}. From this, we observe that the approximated values are very close to the exact ones, for all the four diatomic molecules concerned.  

Using this approximation, we now proceed to solve the radial Schr\"odinger equation by the well-known NU method to obtain the ro-vibrational energy spectrum as given below,  
\beq\label{Energy.En.mp}
	\ba{ll}
	E_{n,\ell,m}
	&=d_4-\f{\a^2\hbar^2}{2\mu}\left(\f{d_3^2}{(n+L)^2}+\f{(n+L)^2}{4}\right),
	\ea 
\eeq
where
\beq
	\ba{ll}
	L&=\f{1}{2}+\sqrt{\f{1}{4}+b^2\b^2+\g d_2},~
	d_3=\f{1}{2}(2+b)b\b^2+\f{\g}{2}(d_2-d_1),~
	d_4=D_e+\f{\g\hbar^2\a^2 d_0}{2\mu}+\f{\a^2\hbar^2 d_3}{2\mu},~
	\b^2=\f{2\mu D_e}{\a^2\hbar^2}.
	\ea
\eeq 

\begin{table}[t]
	\centering
	\scalebox{.7}{
		\begin{tabular}{|rrrr|rrrr|rrrrl|rrrr|}\hline
			$n$ & $\ell$& $m$&$\lambda$& ScH & TiH & VH & CrH& ScH & TiH & VH & CrH&Ref.& ScH & TiH & VH & CrH\\\hline
			&&&&\multicolumn{4}{c|}{$\mu_x=-0.03,\mu_y=-0.01,\mu_z=-0.03$}&\multicolumn{5}{c|}{$\mu_x=\mu_y=\mu_z=0$}&\multicolumn{4}{c|}{$\mu_x=0.03,\mu_y=0.01,\mu_z=0.03$}\\\hline
			0 &  0 &  0 &  0 &  0.104857    &  0.0952017    &  0.109291    &  0.109059&  0.104851 &  0.0951951  & 0.109284 &  0.109052&&  0.104900    &  0.0952442    &  0.109336    &  0.109106\\
			0&  0 &  0 &  1 &  0.104858    &  0.0952025    &  0.109292    &  0.10906&  0.104851 &  0.0951951  & 0.109284 &  0.109052&&  0.104907    &  0.0952495    &  0.109343    &  0.109114\\
			0 &  0 &  0 &  $-$0.25 &  0.104857    &  0.0952016    &  0.109291    &  0.109059&  0.104851 &  0.0951951  & 0.109284 &  0.109052&& 0.104898    &  0.0952428    &  0.109335    &  0.109104\\
			1 &  1 &  0 &  0 &  0.310087    &  0.281755    &  0.323204    &  0.321905 &  0.307516 &  0.279207  & 0.320467 &  0.319101&&  0.310281    &  0.281947    &  0.32341    &  0.322116\\
			1 &  1 &  0 &  1 &  0.310655    &  0.282211    &  0.323789    &  0.322596 &  0.307704 &  0.279358  & 0.320661 &  0.319329&\cite{kdsen2013jmc}$^{\dag}$&0.310877    &  0.282425    &  0.324024    &  0.322842\\
			1 &  1 &  0 &  $-$0.25 &  0.309945    &  0.281641    &  0.323057    &  0.321732 &  0.307469 &  0.279170  & 0.320419 &  0.319044&&0.310131    &  0.281827    &  0.323256    &  0.321934\\
			1 &  1 &  1 &  0 &  0.318539    &  0.290129    &  0.3322    &  0.331123&  0.307516 &  0.279207  & 0.320467 &  0.319101&&   0.319338    &  0.29092    &  0.33305    &  0.331994\\
			1 &  1 &  1 &  1 &  0.32036    &  0.29159    &  0.334076    &  0.333337&  0.307704 &  0.279358  & 0.320661 &  0.319329&&  0.321277    &  0.292476    &  0.335048    &  0.334353\\
			1 &  1 &  1 &  $-$0.25 &  0.318084    &  0.289763    &  0.331731    &  0.330569&  0.307469 &  0.279170  & 0.320419 &  0.319044&&   0.318853    &  0.290531    &  0.332551    &  0.331404\\
			
			2 &  1 &  0 &  0 &  0.500644    &  0.454579    &  0.521641    &  0.518226 &  0.498172 &  0.452133  & 0.519011 &  0.515538&&  0.50083    &  0.454763    &  0.521839    &  0.518428\\
			2 &  1 &  0 &  1 &  0.501236    &  0.455055    &  0.522251    &  0.518947 &  0.498367 &  0.452290  & 0.519213 &  0.515777&\cite{kdsen2013jmc}$^{\S}$&0.501452    &  0.455264    &  0.52248    &  0.519185\\
			2 &  1 &  0 &  $-$0.25 &  0.500496    &  0.454460    &  0.521489    &  0.518046&  0.498123 &  0.452093  & 0.518961 &  0.515479&&   0.500675    &  0.454638    &  0.521679    &  0.518239\\
			
			2 &  1 &  1 &  0 &  0.508772    &  0.462620    &  0.530288    &  0.527061&  0.498172 &  0.452133  & 0.519011 &  0.515538&&  0.50954    &  0.46338    &  0.531105    &  0.527895\\
			2 &  1 &  1 &  1 &  0.510668    &  0.464147    &  0.532243    &  0.529371&  0.498367 &  0.452290  & 0.519213 &  0.515777& &  0.511559    &  0.465006    &  0.533187    &  0.530357\\
			2 &  1 &  1 &  $-$0.25 &  0.508297    &  0.462238    &  0.529799    &  0.526483&  0.498123 &  0.452093  & 0.518961 &  0.515479&& 0.509035    &  0.462973    &  0.530584    &  0.52728\\\hline
		\end{tabular}}
	\caption{\label{table2.energy} Energy of diatomic molecules in Deng-Fan (mid portion) and Dunkl-Deng-Fan (left and right portions) potential for four diatomic hydrides. Ground and excited states are provided. The reference values of ScH, TiH, VH and CrH are 
	$\cite{kdsen2013jmc}^{\dag}: 0.307704129,\, 0.279358078,\, 0.320660504,\, 0.319328938$;~ 
	$\cite{kdsen2013jmc}^{\S}: 0.498367397,\, 0.452290234,\, 0.519212554,\, 0.515776561$.
	}
\end{table}

Following the procedure laid down in \cite{nath2021, nath2021a, nath2021b, nath2023}, the radial solution can be expressed as, 
\beq
	R_{n,\ell,m}(r)=N_{n,\ell,m}^{(r)}\,s^{\bar{\varepsilon}}(1-s)^{L}\,{}_2F_1\left(-n,n+2\bar{\varepsilon}+2L;2\bar{\varepsilon}+1;s\right),
	\eeq 
where 
\beq
	\bar{\varepsilon}=\sqrt{d_0\eps^{(\theta)}_{\ell,m}+\frac{2\mu (D_e-E_{n,m,\ell})}{\a^2\hbar^2}},
\eeq 
and
\beq
	N_{n,\ell,m}^{(r)}=\left[\f{2\bar{\varepsilon}\a(n+\bar{\varepsilon}+L)\G(n+2\bar{\varepsilon}+1)\G(n+2\bar{\varepsilon}+2L)}{n!(n+L)\G(n+2L)\left[\G(2\bar{\varepsilon}+1)\right]^2}\right]^{\f{1}{2}},
\eeq
signifies the normalization constant \cite{stegun}, derived from, 
\beq
	\ds\int_0^{\infty}\frac{R_{n,\ell,m}^2(r)}{r^{2a}}d\chi_{r}=1,d\chi_{r}=r^{2a}dr. 
\eeq 
Assembling all the results, the complete solution of the Deng-Fan potential is given by, $\psi_{n,\ell,m}^{(s_1,s_2,s_3)}= \frac{R_{n,\ell,m}(r)}{r^a}H_{\ell,m}^{(s_3)}(\theta)G_{m}^{(s_1,s_2)}(\phi)$. 
    
\begin{figure}[t]
	\centering
	\includegraphics[width=18cm,height=12cm]{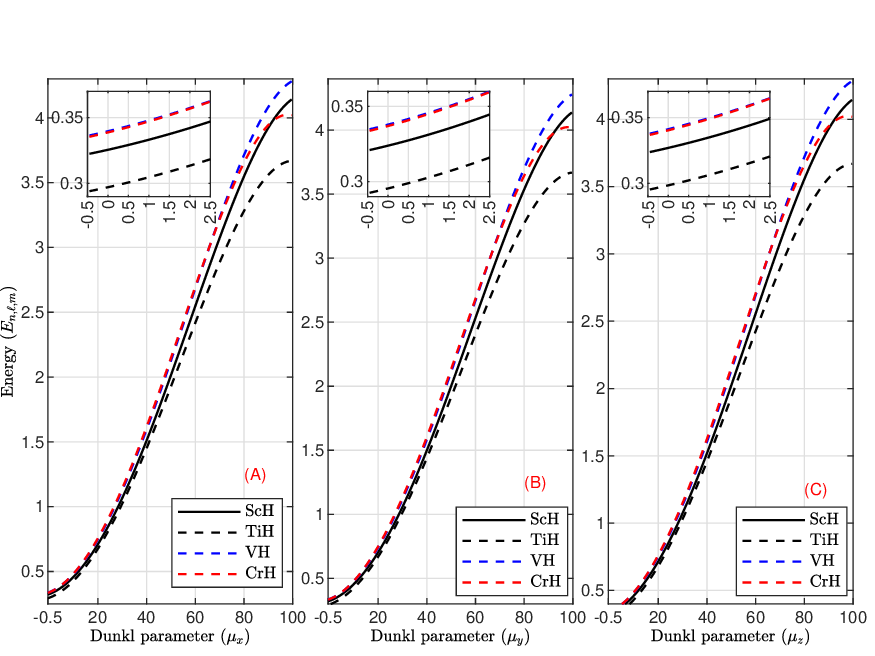}
	\caption{\label{fig2.energy} Comparison of ro-vibrational energy with respect to Dunkl parameters, in $n=\ell=m=1$ state. In (A) $\mu_y=0.75,\mu_z=0.3$, (B) $\mu_x=0.5,\mu_z=0.3$ and (C) $\mu_x=0.5,\mu_y=0.75$. Common parameters are: $s_1=s_2=s_3=1$, $\lambda=-0.25$.}
\end{figure}

\indent
The calculated numerical values of ro-vibrational energies of four representative diatomic molecules, \emph{viz.}, ScH, TiH, VH and CrH, in presence and absence of Dunkl parameters, are produced in Table~\ref{table2.energy}. In all cases, ground and some low-lying excited states are reported, including the $(\ell,m) \neq (0,0)$ states. The left and right segments correspond to $(-)$ve and $(+)$ve Dunkl parameters, while the middle section pertains to the non-Dunkl regime: $\mu_x=\mu_y=\mu_z=0$. The energies seem to be quite stable with changes in $\lambda$ values. In ground state, for all practical purposes, they are almost identical to each other, for all the molecules. For excited states, energies seem to show some deviation from each other as $\lambda$ changes. For the non-Dunkl case, some reference values \cite{kdsen2013jmc} are available, which are duly quoted for comparison. One notices that, when $\mu_x=\mu_y=\mu_z=0$, the Dunkl case results in non-Dunkl energies (from the solution of Schr\"odinger equation). As observed, these reference results corroborate very well with our current results. 
    
The above energies are graphically portrayed in Fig.~\ref{fig2.energy}, with respect to $\mu_x, \mu_y, \mu_z$, for a representative state having quantum numbers, $n=\ell=m=1$. Same four molecules of Table~\ref{table2.energy} are selected. The three panels represent variations of $\mu_x$ keeping $\mu_y, \mu_z$ fixed at 0.75 and 0.3 (A), $\mu_y$ keeping $\mu_x, \mu_z$ fixed at 0.5 and 0.3 (B), and $\mu_z$ for $\mu_x=0.5$, $\mu_y=0.75$. From this figure, it is clear that, in all occasions, energy is a monotone increasing function of Dunkl parameters $\mu_x$, $\mu_y$ and $\mu_z$. 
 
\section{Expectations and Heisenberg uncertainty}\label{sec3.uncertainty}
\begin{figure}[t]
	\centering
	\includegraphics[width=18cm,height=11cm]{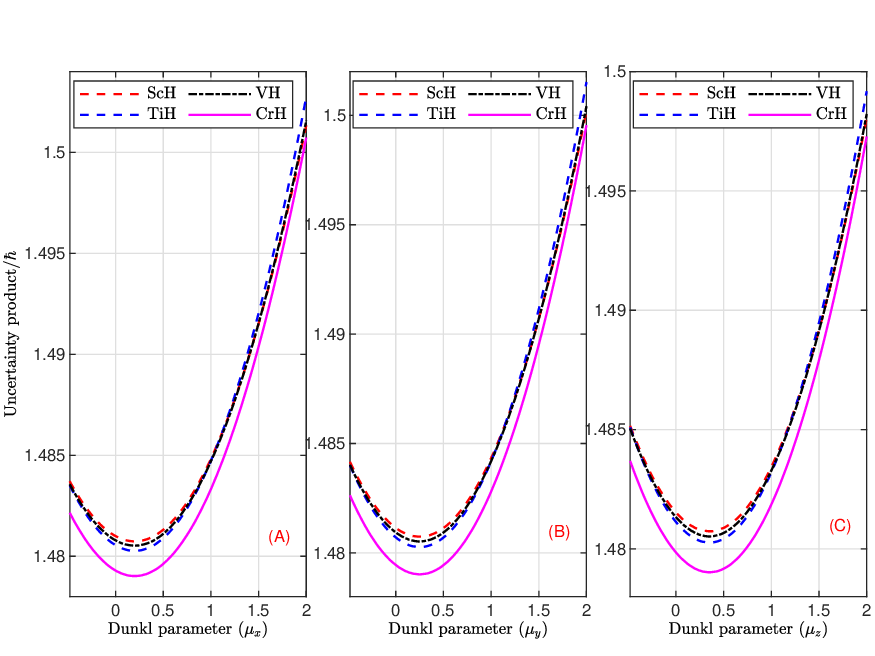}
	\caption{\label{fig3.uncertainty} Radial uncertainty product ${(\Delta r)_{n,\ell,m}(\Delta p_r)_{n,\ell,m}}/{\hbar} $ versus the Dunkl parameters, for four molecules. In (A) $\mu_y=-0.4,\mu_z=-0.3$, (B) $\mu_x=-0.45,\mu_z=-0.3$ and (C) $\mu_x=-0.45,\mu_y=-0.4$. Common parameters are: $s_1=s_2=s_3=1$, $\lambda=0$ and $n=\ell=m=1$.}
\end{figure}

The total density function $\rho_{n,\ell,m}^{(s_1,s_2,s_3)}$ is a product of three independent marginal density functions, $[\frac{R_{n,\ell,m}(r)}{r^{2a}}]^2$, $[H_{\ell,m}^{(s_3)}(\theta)]^2$ and $[G_{m}^{(s_1,s_2)}(\phi)]^2$, which are defined in $r$, $\theta$ and $\phi$ spaces over the domain $r\in[0,\infty)=\mathbb{R}^*$, $\theta\in[0,\pi]$ and $\phi\in[0,2\pi]$. The individual density functions are orthogonal to their respected weighted Lebesgue measures, $d\chi_{r}$, $d\chi_{\theta}$, $d\chi_{\phi}$, and form three weighted $L_2\left(\mathbb{R}^*,d\chi_{r}\right)$, $L_2\left([0,\pi],d\chi_{\theta}\right)$, $L_2\left([0,2\pi],d\chi_{\phi}\right)$ spaces \cite{vinet2013}. Next, the composite (total) density function is orthogonal to the weighted Lebesgue measure product, $d\chi=d\chi_{\phi}d\chi_{\theta}d\chi_{r}$, and the corresponding weighted space is $L_2\left(\mathbb{R}^*\times[0,\pi]\times[0,2\pi],d\chi\right)$. The expectation of a function or operator $\hat{f}(r,\theta,\phi)$ for a given state represented by  $\psi_{n,\ell,m}^{(s_1,s_2,s_3)}(r,\theta,\phi)$, is given as, 
\beq
\left\langle \hat{f}(r,\theta,\phi)\right\rangle_{n,\ell,m}=\ds\int \left[\psi_{n,\ell,m}^{(s_1,s_2,s_3)}(r,\theta,\phi)\right]^*\hat{f}(r,\theta,\phi) \ \psi_{n,\ell,m}^{(s_1,s_2,s_3)}(r,\theta,\phi)d\chi,
\eeq
where $\left[\psi_{n,\ell,m}^{(s_1,s_2,s_3)}(r,\theta,\phi)\right]^*$ is the complex conjugate of $\psi_{n,\ell,m}^{(s_1,s_2,s_3)}(r,\theta,\phi)$. Due to the central nature of the potential, the expectation of an independent operator, $\hat{f}(r,\theta,\phi)=f^{(r)}(r) f^{(\theta)}(\theta) f^{(\phi)}(\phi)$ can be expressed separately as, 
\beq
\ba{lr}
\left\langle \hat{f}(r,\theta,\phi)\right\rangle_{n,\ell,m}&=\ds\int_0^{\infty} \left[\frac{R_{n,\ell,m}(r)}{r^{a}}\right]^*f^{(r)}(r)\frac{R_{n,\ell,m}(r)}{r^{a}}d\chi_r\,\int_0^{\pi} \left[H_{\ell,m}^{(s_3)}(\theta)\right]^*f^{(\theta)}(\theta)H_{\ell,m}^{(s_3)}(\theta)d\chi_{\theta}\\
&\quad\ds\int \left[G_{m}^{(s_1,s_2)}(\phi)\right]^*f^{(\phi)}(\phi)G_{m}^{(s_1,s_2)}(\phi)d\chi_{\phi}.
\ea 
\eeq
Therefore, the expectation $\left\langle r^i\right\rangle_{n,\ell,m}$ can be written as,  
\beq
\ba{ll}
\left\langle r^i\right\rangle_{n,\ell,m}&=\ds\int_0^{\infty} r^iR^2_{n,\ell,m}(r)dr.
\ea  
\eeq
Using the differentiation under the sign of integral, expectation $\left\langle r^i\right\rangle_{n,\ell,m}$ can be expressed as,
\beq
\left\langle r^i\right\rangle_{n,\ell,m}=\ds\frac{(-1)^i[N_{n,\ell,m}^{(r)}]^2}{\a^{i+1}}\sum\limits_{k=0}^{2n}\frac{2\widetilde{B}_{k+2,2}\left(c_0^{(n)},2!c_1^{(n)},\cdots,(k+1)!c_k^{(n)}\right)}{(k+2)!}\left[\lim\limits_{p\rightarrow0}\frac{d^i}{dp^i}\mathcal{B}(2\bar{\varepsilon}+k+p,2L+1)\right].
\eeq 
Furthermore, $\mathcal{B}(a,b)=\frac{\G(a)\G(b)}{\G(a+b)}$ is the Beta function, \(\Gamma(a)\) is the Gamma function, $\Psi^{(d)}(x)=\frac{1}{\G(x)}\frac{d\G(x)}{dx}$ signifies digamma function, whereas 	$\widetilde{B}_{m,\ell}(x_1,x_2,\dots,x_{m-l+1})$ denotes the Bell polynomial defined by \cite{riordan1980}, 
\begin{equation*}
	\widetilde{B}_{m,\ell}(x_1,x_2,\dots,x_{m-l+1})= \sum\limits_{\widehat{\pi}(m,\ell)}\frac{m!}{j_1!j_2!\dots j_{m-\ell+1}!} 
	\left(\frac{x_1}{1!}\right)^{j_1} \left(\frac{x_2}{2!}\right)^{j_2} \dots  \bigg( \frac{x_{m-\ell+1}}{(m-\ell+1)!} \bigg)^{j_{m-\ell+1}}.
\end{equation*} 
Here $\widehat{\pi}(m,\ell)$ refers to the set of partitions, such that, $\ds j_1+j_2+\dots+j_{m-\ell+1}=\ell,~j_1+2j_2+\dots+(m-\ell+1)j_{m-\ell+1}=m$, and
\beq
c_i^{(n)}=\left\{\ba{ll}\frac{(-n)_i(n+2\bar{\varepsilon}+2L)_i}{(2\bar{\varepsilon}+1)_ii!},&i\le n\\0,&i>n\ea\right\},
\eeq 
\((a)_i= a(a+1)...(a+i-1)\) signifies the Pochhammer symbol.
For $i=1,2$, the following expectations can be written, 
\beq
\ba{ll}
\left\langle r\right\rangle_{n,\ell,m}&=\ds-\frac{[N_{n,\ell,m}^{(r)}]^2}{\a^{2}}\sum\limits_{k=0}^{2n}\frac{2\widetilde{B}_{k+2,2}\left(c_0^{(n)},2!c_1^{(n)},\cdots,(k+1)!c_k^{(n)}\right)}{(k+2)!}\mathcal{B}(2\bar{\varepsilon}+k,2L+1)\left[\Psi^{(d)}(2\bar{\varepsilon}+k)-\Psi^{(d)}(2\bar{\varepsilon}+2L+k+1)\right],\\
\left\langle r^2\right\rangle_{n,\ell,m}&=\ds\frac{[N_{n,\ell,m}^{(r)}]^2}{\a^{3}}\sum\limits_{k=0}^{2n}\frac{2\widetilde{B}_{k+2,2}\left(c_0^{(n)},2!c_1^{(n)},\cdots,(k+1)!c_k^{(n)}\right)}{(k+2)!}\mathcal{B}(2\bar{\varepsilon}+k,2L+1)\left[\Psi^{(p)}(1,2\bar{\varepsilon}+k)-\Psi^{(p)}(1,2\bar{\varepsilon}+2L+k+1)\right.\\
&\quad\left. \left\{\Psi^{(d)}(2\bar{\varepsilon}+k)-\Psi^{(d)}(2\bar{\varepsilon}+2L+k+1)\right\}^2\right],
\ea
\eeq
where $\Psi^{(p)}(n,x)$ represents the Polygamma function, defined by $\Psi^{(p)}(n,x)=\frac{d^n}{dx^n}\Psi^{(d)}(x)$, and $\Psi^{(p)}(0,x)=\Psi^{(d)}(x)$. Next, we provide two important expectations of inverse power, i.e., $r^{-2}$ and $r^{-1}$, which are found to be, 
\beq
\ba{ll}
\left\langle \frac{1}{r^2}\right\rangle_{n,\ell,m}&=\ds\int_0^{\infty} \frac{1}{r^2}R^2_{n,\ell,m}(r)dr=\ds\a [N_{n,\ell,m}^{(r)}]^2\sum\limits_{k=0}^{2n}\frac{2\widetilde{B}_{k+2,2}\left(c_0^{(n)},2!c_1^{(n)},\cdots,(k+1)!c_k^{(n)}\right)}{(k+2)!}\int_0^1s^{2\bar{\varepsilon}+k-1}(1-s)^{2L}\frac{ds}{(\ln[s])^2}\\
&=\ds\a[N_{n,\ell,m}^{(r)}]^2\sum\limits_{k=0}^{2n}\frac{\widetilde{B}_{k+2,2}\left(c_0^{(n)},2!c_1^{(n)},\cdots,(k+1)!c_k^{(n)}\right)}{(k+2)!}\sum\limits_{i=0}^{\lfloor\rfloor}(-1)^i\binom{2L}{i}(2\bar{\varepsilon}+k+i)^2\ln(2\bar{\varepsilon}+k+i),
\ea
\eeq
where $\sum\limits_{i=0}^{\lfloor\rfloor}$ is a finite series, if $2L\in\mathbb{N}$;  otherwise it is an infinite series. The expectation $\left\langle \frac{1}{r}\right\rangle$ is given as, 
\beq
\ba{ll}
\left\langle \frac{1}{r}\right\rangle_{n,\ell,m}&=\ds\int_0^{\infty} \frac{1}{r}R^2_{n,\ell,m}(r)dr=\ds-[N_{n,\ell,m}^{(r)}]^2\sum\limits_{k=0}^{2n}\frac{2\widetilde{B}_{k+2,2}\left(c_0^{(n)},2!c_1^{(n)},\cdots,(k+1)!c_k^{(n)}\right)}{(k+2)!}\int_0^1s^{2\bar{\varepsilon}+k-1}(1-s)^{2L}\frac{ds}{\ln[s]},\\
&=\ds-[N_{n,\ell,m}^{(r)}]^2\sum\limits_{k=0}^{2n}\frac{2\widetilde{B}_{k+2,2}\left(c_0^{(n)},2!c_1^{(n)},\cdots,(k+1)!c_k^{(n)}\right)}{(k+2)!}\sum\limits_{i=0}^{2L}(-1)^{i-1}\ln[\G(2L-i+2\bar{\varepsilon}+k)], ~\mbox{if}~2L\in\mathbb{N}.
\ea
\eeq 
Note that, for $2L\notin\mathbb{N}$, it needs to be pursued numerically. 
Now, one can find the root mean square (RMS) $\left(\Delta r\right)_{n,\ell,m}=\sqrt{\left\langle r^2\right\rangle_{n,\ell,m}-\left\langle r\right\rangle_{n,\ell,m}^2}$
of $r$.
The expectation of the potential function, $V(r)$, is given as,   
\beq
\ba{lr}
\left\langle V(r)\right\rangle_{n,\ell,m}&=\ds D_e+\frac{b^2D_e[N_{n,\ell,m}^{(r)}]^2}{\a}\sum\limits_{i=0}^{2n}\frac{2\widetilde{B}_{i+2,2}\left(c_0^{(n)},2!c_1^{(n)},\cdots,(i+1)!c_i^{(n)}\right)}{(i+2)!}\mathcal{B}(2\bar{\varepsilon}+i+2,2L-1)\\
&\quad\ds-\frac{2bD_e[N_{n,\ell,m}^{(r)}]^2}{\a}\sum\limits_{i=0}^{2n}\frac{2\widetilde{B}_{i+2,2}\left(c_0^{(n)},2!c_1^{(n)},\cdots,(i+1)!c_i^{(n)}\right)}{(i+2)!}\mathcal{B}(2\bar{\varepsilon}+i+1,2L).
\ea
\eeq
The radial momentum operator, $r_r$, in Schr\"odinger-Dunkl system is defined by $p_r=\ds-i\hbar\left(\frac{\partial}{\partial r}+\frac{a}{r}\right)$, satisfying the commutation relation $[r,p_r]=i\hbar$ and $p_r^2=-\hbar^2\left(M_r+\frac{a(a-1)}{r^2}\right)$.
The expectation of $p_r$ \cite{mom.op}, can be expressed as,
\beq
\ba{ll}
\left\langle p_r\right\rangle_{n,\ell,m} &=0,\\
\left\langle p_r^2\right\rangle_{n,\ell,m}&
=2\mu E_{n,\ell,m}+\hbar^2\g\left\langle\frac{1}{r^2 }\right\rangle_{n,\ell,m}-2\mu\left\langle V\right\rangle_{n,\ell,m}.
\ea
\eeq
Then the RMS of radial momentum operator can be expressed as, 
\beq
\left(\Delta p_r\right)_{n,\ell,m}=\sqrt{\left\langle p_r^2\right\rangle_{n,\ell,m}-\left\langle p_r\right\rangle_{n,\ell,m}^2}=\sqrt{\left\langle p_r^2\right\rangle_{n,\ell,m}}.
\eeq
Thus, the product of RMS of $r$ and $p$ is found to satisfy the following Heisenberg uncertainty relation \cite{Heisenberg,Kennard,dehesa.uncertainty}, 
\beq
(\Delta r)_{n,\ell,m}(\Delta p_r)_{n,\ell,m}\ge \frac{\hbar}{2}.
\eeq 
In Fig.~\ref{fig3.uncertainty}, we have plotted radial uncertainty product ${(\Delta r)_{n,\ell,m}(\Delta p_r)_{n,\ell,m}}/{\hbar}$ against $\mu_x, \mu_y, \mu_z$ for four molecular hydrides. The three panels from left to right correspond to variations with respect to $\mu_x$ (or $\mu_y, \mu_z$) keeping the other two fixed. In all cases, we notice that the products monotonically increase as Dunkl parameter enhances. At smaller values of $\mu$'s the plots for all molecules remain practically coincident, showing a gradual tendency to show some separation as the same rises. It is observed that the uncertainty product satisfies the Heisenberg uncertainty principle. 

\section{Shannon entropy of Deng-Fan potential in Dunkl framework in position space}\label{sec4.shannon}
For an arbitrary state, $\psi_{n,\ell,m}^{(s_1,s_2,s_3)}$, the relevant $S$, defined in $L_2\left(\mathbb{R}^*\times[0,\pi]\times[0,2\pi],d\chi\right)$ space, is given by \cite{shannon},  
\beq
	\ba{ll}
	\mathcal{S}(\rho_{n,\ell,m}^{(s_1,s_2,s_3)})&=-\ds\int|\psi_{n,\ell,m}^{(s_1,s_2,s_3)}(r,\theta,\phi)|^2\ln[|\psi_{n,\ell,m}^{(s_1,s_2,s_3)}(r,\theta,\phi)|^2]d\chi 
	=\mathcal{S}_{n,\ell,m}^{(r)}+\mathcal{S}_{\ell,m}^{(\theta)}+\mathcal{S}_m^{(\phi)}, 
	\ea 
\eeq 
where $\mathcal{S}_{n,\ell,m}^{(r)}$, $\mathcal{S}_{n,\ell,m}^{(\theta)}$ and $\mathcal{S}_{n,\ell,m}^{(\phi)}$ are Shannon entropies of the marginal density functions in $r, \theta, \phi$ spaces respectively. In the following, the marginal Shannon entropies in $r, \theta$ and $\phi$ coordinates are offered one by one. 

\subsection{Shannon entropy of the marginal density function $\left[\frac{R_{n,\ell,m}(r)}{r^{a}}\right]^2$}
The entropy  $\mathcal{S}_{n,\ell,m}^{(r)}$ corresponding to the radial density function is defined by: 
\beq
	\ba{ll}\label{sha.r.nu}
	\mathcal{S}_{n,\ell,m}^{(r)}&=-\ds\int_0^{\infty} \frac{R_{n,\ell,m}^2(r)}{r^{2a}}\ln\left[\frac{R_{n,\ell,m}^2(r)}{r^{2a}}\right]d\chi_{r}.
	\ea
\eeq
This can be further recast as,  
\beq
	\ba{ll}\label{sha.r.qa}
	\mathcal{S}_{n,\ell,m}^{(r)}&=\mathcal{S}_{n,\ell,m}^{(r,1)}-\ln[(N_{n,\ell,m}^{(r)})^2]+2(N_{n,\ell,m}^{(r)})^2\left[\bar{\varepsilon}\,\mathcal{S}^{(r,2)}+L\,\mathcal{S}^{(r,3)}\right]+2a\left\langle \ln[r]\right\rangle_{n,\ell,m},
	\ea
\eeq
where
\beq
	\ba{ll}\label{sha.r1}
	\mathcal{S}_{n,\ell,m}^{(r,1)}&=-\ds\frac{(N_{n,\ell,m}^{(r)})^2}{\a}\int_0^1s^{2\bar{\varepsilon}-1}(1-s)^{2L}\left[{}_2F_1\left(-n,n+2\bar{\varepsilon}+2L;2\bar{\varepsilon}+1;s\right)\right]^2\ln\left(\left[{}_2F_1\left(-n,n+2\bar{\varepsilon}+2L;2\bar{\varepsilon}+1;s\right)\right]^2\right)ds,
	\ea
\eeq
\beq\ba{ll}\label{sha.r2}
	\mathcal{S}^{(r,2)}&=-\ds\frac{1}{\a}\int_0^1s^{2\bar{\varepsilon}-1}(1-s)^{2L}\left[{}_2F_1\left(-n,n+2\bar{\varepsilon}+2L;2\bar{\varepsilon}+1;s\right)\right]^2\ln[s]ds\\
	&=-\ds\frac{1}{\a}\sum\limits_{i=0}^{2n}\frac{2\widetilde{B}_{i+2,2}\left(c_0^{(n)},2!c_1^{(n)},\cdots,(i+1)!c_i^{(n)}\right)}{(i+2)!}\mathcal{B}(2\bar{\varepsilon}+i,2L+1)\left[\Psi^{(d)}(2\bar{\varepsilon}+i)-\Psi^{(d)}(2\bar{\varepsilon}+2L+i+1)\right],
	\ea
\eeq
\beq
	\ba{ll}\label{sha.r3}
	\mathcal{S}^{(r,3)}&=-\ds\frac{1}{\a}\int_0^1s^{2\bar{\varepsilon}-1}(1-s)^{2L}\left[{}_2F_1\left(-n,n+2\bar{\varepsilon}+2L;2\bar{\varepsilon}+1;s\right)\right]^2\ln[1-s]ds\\
	&=-\ds\frac{1}{\a}\sum\limits_{i=0}^{2n}\frac{2\widetilde{B}_{i+2,2}\left(c_0^{(n)},2!c_1^{(n)},\cdots,(i+1)!c_i^{(n)}\right)}{(i+2)!}\mathcal{B}(2\bar{\varepsilon}+i,2L+1)\left[\Psi^{(d)}(2L+1)-\Psi^{(d)}(2\bar{\varepsilon}+2L+i+1)\right].
	\ea 
\eeq
 Following \cite{bi,gw,integrals}, the expectation $\left\langle \ln[r]\right\rangle_{n,\ell,m}$ can be simplified as:
\beq
	\ba{ll}\label{j.j.r.l}
	\left\langle \ln[r]\right\rangle_{n,\ell,m}&=\ds\int_0^{\infty} [R_{n,\ell,m}^2]\ln[r]dr\\
	&=\ds\ln[\a]+\frac{[N_{n\ell,m}^{(r)}]^2}{\a}\sum\limits_{i=0}^{2n}\sum\limits_{j=0}^{\lfloor2L\rfloor}\frac{2\widetilde{B}_{i+2,2}\left(c_0^{(n)},2!c_1^{(n)},\cdots,(i+1)!c_i^{(n)}\right)}{(i+2)!}(-1)^j\binom{2L}{j}\int_0^1s^{2\bar{\varepsilon}+i+j-1}\ln\left[\ln\left(\frac{1}{s}\right)\right]ds\\
	&=\ds\ln[\a]+\frac{[N_{n\ell,m}^{(r)}]^2}{\a}\sum\limits_{i=0}^{2n}\sum\limits_{j=0}^{\lfloor2L\rfloor}\frac{2\widetilde{B}_{i+2,2}\left(c_0^{(n)},2!c_1^{(n)},\cdots,(i+1)!c_i^{(n)}\right)}{(i+2)!}(-1)^j\binom{2L}{j}\frac{\g^{(E)}+\ln[2\bar{\varepsilon}+i+j]}{2\bar{\varepsilon}+i+j},
	\ea 
\eeq
where $\g^{(E)}=\ds\lim_{n\rightarrow \infty}\left(\sum\limits_{k=1}^n\frac{1}{k}-\ln[n]\right)$ is the Euler's constant. An alternative form of Euler's constant $\g^{(E)}$ can be written as \cite{euler} $\g^{(E)}=\ds\int_1^{\infty}\left\{\frac{1}{\lfloor x\rfloor}-\frac{1}{x}\right\}dx$, where $\lfloor x\rfloor$ is the floor function, representing the greatest integer, not greater than $x$, i.e., $\lfloor x\rfloor\le x$. The numerical value of the Euler's constant up to 10 decimal places is $\g^{(E)}\approx0.5772156649$. 

\begin{table}
	\scalebox{.65}{
		\begin{tabular}{|rrrr|rrrrl|rrrrl|rrrrl|}\hline
			$n$& $\ell$& $m$&$\lambda$&$\mathcal{S}^{(r)}$ & $\mathcal{S}^{(\theta)}$ & $\mathcal{S}^{(\phi)}$ & $\mathcal{S}$&$\eps_{\%}(\mathcal{S})$&$\mathcal{S}^{(r)}$ & $\mathcal{S}^{(\theta)}$ & $\mathcal{S}^{(\phi)}$ & $\mathcal{S}$&$\eps_{\%}(\mathcal{S}^{(r)})$&$\mathcal{S}^{(r)}$ & $\mathcal{S}^{(\theta)}$ & $\mathcal{S}^{(\phi)}$ & $\mathcal{S}$&$\eps_{\%}(\mathcal{S})$\\\hline
			&&&&\multicolumn{5}{c|}{$\mu_x=-0.03,\mu_y=-0.01,\mu_z=-0.03$}&\multicolumn{5}{c|}{$\mu_x=\mu_y=\mu_z=0$}&\multicolumn{5}{c|}{$\mu_x=0.03,\mu_y=0.01,\mu_z=0.03$}\\\hline
			0  & 0 &  0  & 0 &  0.278273 &  0.776867  & 1.89453 &  2.94967 &0.00001&  0.360923 &  0.693147  & 1.83788 &  2.89195&0.000000007& 0.443589 &  0.609794  & 1.78353 &  2.83691 &0.000000003\\
			0  & 0 &  0  & 1 &  0.278272 &  0.776867  & 1.89453 &  2.94967 &0.00001&  0.360923 &  0.693147  & 1.83788 &  2.89195&0.000000007&  0.443587 &  0.609794  & 1.78353 &  2.83691 &0.000000003\\
			0  & 0 &  0  & -0.25 &  0.278273 &  0.776867  & 1.89453 &  2.94967&0.00001&  0.360923 &  0.693147  & 1.83788 &  2.89195&0.000000007&  0.443589 &  0.609794  & 1.78353 &  2.83691 &0.0000000005\\
			
			1  & 1 &  0  & 0 &  0.620275 &  0.312826  & 1.89453 &  2.82763&0.0001&  0.706035 &  0.261202  & 1.83788 &  2.80511&0.000002& 0.793246 &  0.120157  & 1.78353 &  2.69693 &0.00003\\
			1  & 1 &  0  & 1 &  0.620165 &  0.312826  & 1.89453 &  2.82752 &0.0001&  0.705998 &  0.261202  & 1.83788 &  2.80508&0.00000005&  0.793123 &  0.120157  & 1.78353 &  2.69681 &0.00003\\
			1  & 1 &  0  & -0.25 &  0.620303 &  0.312826  & 1.89453 &  2.82766&0.0001 & 0.706045 &  0.261202  & 1.83788 &  2.80512&0.0000003&   0.793277 &  0.120157  & 1.78353 &  2.69696 &0.00003\\
			1  & 1 &  1  & 0 &  0.622477 &  0.445102  & 1.5944 &  2.66198 &0.0004&  0.706035 &  0.568054  & 1.83788 &  3.11197&0.000002&  0.79581 &  0.291156  & 1.46928 &  2.55625 &0.00008\\
			1  & 1 &  1  & 1 &  0.622124 &  0.445102  & 1.5944 &  2.66163 &0.0004&  0.705998 &  0.568054  & 1.83788 &  3.11193&0.00000005& 0.795407 &  0.291156  & 1.46928 &  2.55585&0.00009\\
			1  & 1 &  1  & -0.25 &  0.622565 &  0.445102  & 1.5944 &  2.66207 &0.0004&  0.706045 &  0.568054  & 1.83788 &  3.11198&0.0000003&  0.795911 &  0.291156  & 1.46928 &  2.55635&0.00008\\
			2  & 1 &  0  & 0 &  0.847638 &  0.312826  & 1.89453 &  3.055 &0.0002&  0.937246 &  0.261202  & 1.83788 &  3.03632&0.00007& 1.02836 &  0.120157  & 1.78353 &  2.93204 &0.00005 \\
			2  & 1 &  0  & 1 &  0.847522 &  0.312826  & 1.89453 &  3.05488 &0.0002&  0.937206 &  0.261202  & 1.83788 &  3.03628&0.00002&  1.02823 &  0.120157  & 1.78353 &  2.93191 &0.00004\\
			2  & 1 &  0  & -0.25 &  0.847667 &  0.312826  & 1.89453 &  3.05503 &0.0002&  0.937256 &  0.261202  & 1.83788 &  3.03633&0.00007& 1.02839 &  0.120157  & 1.78353 &  2.93208 &0.00004\\
			2  & 1 &  1  & 0 &  0.849923 &  0.445102  & 1.5944 &  2.88943 &0.0006 &  0.937246 &  0.568054  & 1.83788 &  3.34318&0.00007&   1.03102 &  0.291156  & 1.46928 &  2.79146 &0.0001\\
			2  & 1 &  1  & 1 &  0.849551 &  0.445102  & 1.5944 &  2.88906 &0.0006 &  0.937206 &  0.568054  & 1.83788 &  3.34314&0.00002& 1.03059 &  0.291156  & 1.46928 &  2.79103 &0.0001\\
			2  & 1 &  1  & -0.25 &  0.850017 &  0.445102  & 1.5944 &  2.88952 &0.0004  &  0.937256 &  0.568054  & 1.83788 &  3.34319&0.00007&  1.03113 &  0.291156  & 1.46928 &  2.79157&0.0001\\\hline
	\end{tabular}}
	\caption{\label{table3.tih}$\mathcal{S}^{(r)}, \mathcal{S}^{(\theta)}, \mathcal{S}^{(\phi)}$ and $\mathcal{S}$ of TiH molecule in Deng-Fan potential in the Dunkl-Schr\"odinger framework, in ground and some low-lying excited states. The $(-)$ve and $(+)$ve Dunkl parameters are in the left and right segments, while middle segment refers to non-Dunkl case. In all cases, three different $\lambda$ considered. For details, see text.}
\end{table}

\subsubsection{Factorization method for $\mathcal{S}_{n,\ell,m}^{(r,1)}$}
The expression for $\mathcal{S}_{n,\ell,m}^{(r,1)}$, in Eq.~(\ref{sha.r1}) can be further simplified from a consideration of the hyper-geometric function  ${}_2F_1\left(-n,n+2\bar{\varepsilon}+2L;2\bar{\varepsilon}+1;s\right)$, as follows,  
\beq
{}_2F_1\left(-n,n+2\bar{\varepsilon}+2L;2\bar{\varepsilon}+1;s\right)=a^{(r)}\prod\limits_{i=1}^n(s-s_i),
\eeq 
where $a^{(r)}_n=\frac{(-n)_n(n+2\bar{\varepsilon}+2L)_n}{(2\bar{\varepsilon}+1)_nn!}$ is the leading term of ${}_2F_1\left(-n,n+2\bar{\varepsilon}+2L;2\bar{\varepsilon}+1;s\right)$ and $s_j(j=1,2,\cdots,n)$ are roots of the equation  ${}_2F_1\left(-n,n+2\bar{\varepsilon}+2L;2\bar{\varepsilon}+1;s\right)=0$, such that $0\le s_1\le s_2\le s_3\le\cdots\le s_n\le 1$. Then the integral in $\mathcal{S}^{(r,1)}_{n,\ell,m}$ can be expressed as \cite{log.pot,lop.pot2,log.pot3.dn,log.pot.td.ho,tog.pot.makarov}: 
\beq
\ba{lr}\label{sha.r1.sum}
\mathcal{S}_{n,\ell,m}^{(r,1)}&=\ds(N_{n,\ell,m}^{(r)})^2\sum\limits_{j=1}^n\sum\limits_{i=1}^{2n}\frac{2\widetilde{B}_{i+2,2}\left(c_0^{(n)},2!c_1^{(n)},\cdots,(i+1)!c_i^{(n)}\right)\ln\left(\frac{1-s_j}{s_j}\right)^2}{(i+2)!\,\a}\mathcal{B}(s_j;2\bar{\varepsilon}+i,2L+1)\\
&-\ds\frac{(N_{n,\ell,m}^{(r)})^2}{\a}\sum\limits_{j=1}^n\left[\mathcal{J}_{i,j}^{(r,R)}+\mathcal{J}_{i,j}^{(r,L)}\right]-\ds\ln\left[a^{(r)}_n\prod\limits_{j=1}^n\left(1-s_j\right)\right]^2\\
\ea
\eeq
where
\beq\label{jr}
\mathcal{J}_{j}^{(r,R)}=\int_0^{s_j}s^{2\bar{\varepsilon}-1}(1-s)^{2L}\left\{a_n^{(r)}\prod\limits_{i=1,i\ne j}^n(s-s_i)\right\}^2(s-s_j)^2\ln\left(1-\frac{s}{s_j}\right)^2ds,
\eeq 
\beq\label{jl}
\mathcal{J}_{j}^{(r,L)}=\int_{s_j}^1s^{2\bar{\varepsilon}-1}(1-s)^{2L}\left\{a_n^{(r)}\prod\limits_{i=1,i\ne j}^n(s-s_i)\right\}^2(s-s_j)^2\ln\left(\frac{s-s_j}{1-s_j}\right)^2ds,
\eeq 
are improper integrals and $\mathcal{B}(s_j;2\bar{\varepsilon}+i,2L+1)=\ds\int_0^{s_j}s^{2\bar{\varepsilon}+i-1}(1-s)^{2L}ds$. The incomplete beta function $\mathcal{B}(z;a,b)$ reduces to the complete beta function $\mathcal{B}(a,b)$, if $z=0$, i.e., $\mathcal{B}(0;a,b)=\mathcal{B}(a,b)$. 
From the integrals in Eq.~(\ref{jr}), it is clear that $s_j$ is the only point of infinite discontinuity such that, 
\beq
\ba{ll}
\ds-(1-s)^{2L}\left\{a_n^{(r)}\prod\limits_{i=1,i\ne j}^n(s-s_i)\right\}^2(s-s_j)^2\ln\left(1-\frac{s}{s_j}\right)^2>0,~\mbox{for~all~}s\in(0,s_j),\\
\ds\lim_{s\rightarrow s_j-}(1-s)^{2L}\left\{a_n^{(r)}\prod\limits_{i=1,i\ne j}^n(s-s_i)\right\}^2(s-s_j)^2\ln\left(1-\frac{s}{s_j}\right)^2=0,
\ea 
\eeq 
and the product function $-(1-s)^{2L}\left\{a_n^{(r)}\prod\limits_{i=1,i\ne j}^n(s-s_i)\right\}^2(s-s_j)^2\ln\left(1-\frac{s}{s_j}\right)^2$ is continuous in $(0,s_j)$. Therefore, the integral $\mathcal{J}_{j}^{(r,R)}$ exists \cite{improper,log.pot3.dn}. To find $\mathcal{J}_{j}^{(r,R)}$, we express this integral in an alternative form by using a transformation $s=s_j(1-x)$ as follows, 
\beq
\ba{ll}\label{j.j.r.r}
\mathcal{J}_{j}^{(r,R)}
&=\ds\sum\limits_{i=1}^{2n}\sum\limits_{k=0}^{\lfloor2L\rfloor}\frac{4\widetilde{B}_{i+2,2}\left(c_0^{(n)},2!c_1^{(n)},\cdots,(i+1)!c_i^{(n)}\right)}{(i+2)!}s_j^{2\bar{\varepsilon}+i+k}(1-s_j)^{2L-k}\binom{2L}{k}\int_0^{1}(1-x)^{2\bar{\varepsilon}+i-1}x^k\ln x\,dx\\
&=\ds\sum\limits_{i=1}^{2n}\sum\limits_{k=0}^{\lfloor 2L\rfloor}\left\{\frac{4\widetilde{B}_{i+2,2}\left(c_0^{(n)},2!c_1^{(n)},\cdots,(i+1)!c_i^{(n)}\right)}{(i+2)!}s_j^{2\bar{\varepsilon}+i+k}(1-s_j)^{2L-k}\binom{2L}{k}\right\}\\
&\quad\times \mathcal{B}(2\bar{\varepsilon}+i,k+1)[\Psi^{(d)}(k+1)-\Psi^{(d)}(2\bar{\varepsilon}+i+k+1)].
\ea 
\eeq  
Note that, $\sum\limits_{k=0}^{\lfloor2L\rfloor}$ is finite in Eqs.~(\ref{j.j.r.l}),(\ref{j.j.r.r}) if $2L$ belong to $\mathbb{N}$, otherwise it is a convergent infinite series.
Similarly, we can say that the integral $\mathcal{J}_{j}^{(r,L)}$ exists \cite{improper,log.pot3.dn}. To find $\mathcal{J}_{j}^{(r,L)}$, we express this integral in an alternative form by using another transformation of the form, such that $s=s_j+(1-s_j)y$, 
\beq
\ba{ll}
\mathcal{J}_{j}^{(r,L)}
&=\ds\sum\limits_{i=1}^{2n}\sum\limits_{k=0}^{\lfloor2\bar{\varepsilon}+i-1\rfloor}\frac{4\widetilde{B}_{i+2,2}\left(c_0^{(n)},2!c_1^{(n)},\cdots,(i+1)!c_i^{(n)}\right)}{(i+2)!}s_j^{2\bar{\varepsilon}+i-k-1}(1-s_j)^{2L+k-1}\binom{2\bar{\varepsilon}+i-1}{k}\int_0^{1}(1-y)^{2L}y^k\ln y\,dy\\
&=\ds\sum\limits_{i=1}^{2n}\sum\limits_{k=0}^{\lfloor2\bar{\varepsilon}+i-1\rfloor}\left\{\frac{4\widetilde{B}_{i+2,2}\left(c_0^{(n)},2!c_1^{(n)},\cdots,(i+1)!c_i^{(n)}\right)}{(i+2)!}s_j^{2\bar{\varepsilon}+i-k-1}(1-s_j)^{2L+k-1}\binom{2\bar{\varepsilon}+i-1}{k}\right\}\\
&\quad\times \mathcal{B}(k+1,2L+1)[\Psi^{(d)}(k+1)-\Psi^{(d)}(2L+k+2)].
\ea 
\eeq 
Note that, $\sum\limits_{k=0}^{\lfloor2\bar{\varepsilon}+i-1\rfloor}$ is finite if $2\bar{\varepsilon}+i-1\in\mathbb{N}$; otherwise it is an infinite series and convergent. This completes the radial Shannon entropy function.

\begin{figure}[t]
	\centering
	\includegraphics[width=18cm,height=10cm]{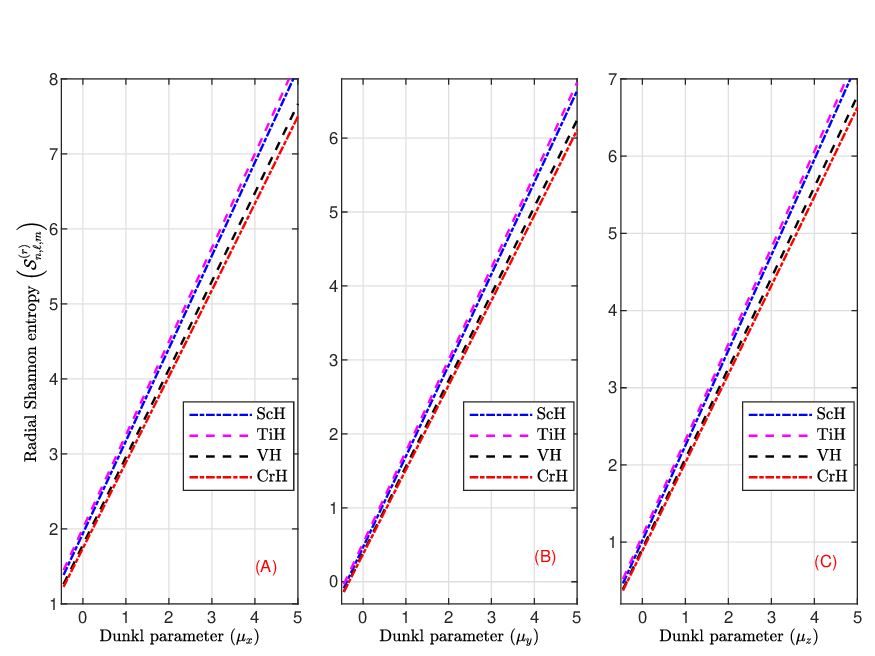}
	\caption{\label{fig3.shannon-r} $\mathcal{S}^{(r)}_{n,\ell,m}$ of radial density functions $\left[\frac{R_{n,\ell,m}(r)}{r^{a}}\right]^2$ of molecules in Deng-Fan potential in Dunkl framework. In (A) $\mu_y=0.75$, $\mu_z=0.3$ (B) $\mu_x=-0.45$, $\mu_z=0.3$ and (C) $\mu_x=-0.45$, $\mu_y=0.75$. All correspond to $n=\ell=m=1$ state, for $\lambda=0$. } 
\end{figure}

\subsection{Shannon entropy of the marginal density function $\left[H_{\ell,m}^{(s_3)}(\theta)\right]^2$}
\noindent Now, $\mathcal{S}_{\ell,m}^{(\theta)}$ of the angular density function $\left[H_{\ell,m}^{(s_3)}(\theta)\right]^2$ is given by, 
\beq\label{sha.theta.nu}
	\mathcal{S}_{\ell,m}^{(\theta)}=-\ds\int_{0}^{\pi}\left[H_{\ell,m}^{(s_3)}(\theta)\right]^2\ln[\left[H_{\ell,m}^{(s_3)}(\theta)\right]^2]d\chi_{\theta}.
	\eeq
	This can be expressed as follows, 
	\beq\label{sha.theta.qa}
	\mathcal{S}_{\ell,m}^{(\theta)}=\mathcal{S}_{\ell,m}^{(\theta,1)}-\ln[(N_{\ell,m}^{(\theta)})^2]+2[N_{\ell,m}^{(\theta)}]^2\left[l_3\mathcal{S}^{(\theta,2)}+l_4\mathcal{S}^{(\theta,3)}\right],
\eeq
where
\beq\label{sha.theta1}
\ba{ll}
	\mathcal{S}_{\ell,m}^{(\theta,1)}&=\ds-\frac{(N_{\ell,m}^{(\theta)})^2}{2^{a-\frac{1}{2}}}\ds\int_{-1}^{1}(1-q)^{\mu_x+\mu_y+2l_3}(1+q)^{\mu_z+2l_4-\frac{1}{2}}\left[P_{\ell}^{(\a_2,\b_2)}(q)\right]^2\ln\left(\left[P_{\ell}^{(\a_2,\b_2)}(q)\right]^2\right)dq,
	\ea 
\eeq
\beq
	\ba{ll}\label{sha.theta2}
	\mathcal{S}^{(\theta,2)}&=\ds-\frac{1}{2^{a-\frac{1}{2}}}\ds\int_{-1}^{1}(1-q)^{\mu_x+\mu_y+2l_3}(1+q)^{\mu_z+2l_4-\frac{1}{2}}\left[P_{\ell}^{(\a_2,\b_2)}(q)\right]^2\ln\left[1-q\right]dq\\
	&=\ds-\frac{2[(\a_2+1)_{\ell}]^2}{[\ell!]^2}\sum\limits_{i=0}^{2\ell}\frac{2^{2l_3+2l_4+i}\widetilde{B}_{i+2,2}\left(c_0^{(\ell)},2!c_1^{(\ell)},\cdots,(i+1)!c_i^{(\ell)}\right)}{(i+2)!}\mathcal{B}\left(\mu_x+\mu_y+2l_3+i+1,\mu_z+2l_4+\frac{1}{2}\right)\\
	&\quad\hspace{5cm}\ds\times\left[\ln2+\Psi^{(d)}\left(\mu_x+\mu_y+2l_3+i+1\right)-\Psi^{(d)}\left(a+2l_3+2l_4+\frac{1}{2}+i\right)\right],
	\ea
\eeq
\beq\label{sha.theta3}
	\ba{ll}
	\mathcal{S}^{(\theta,3)}&=-\frac{1}{2^{a-\frac{1}{2}}}\ds\int_{-1}^{1}(1-q)^{\mu_x+\mu_y+2l_3}(1+q)^{\mu_z+2l_4-\frac{1}{2}}\left[P_{\ell}^{(\a_2,\b_2)}(q)\right]^2\ln\left[1+q\right]dq\\
	&=\ds-\frac{2[(\a_2+1)_{\ell}]^2}{[\ell!]^2}\sum\limits_{i=0}^{2\ell}\frac{2^{2l_3+2l_4+i}\widetilde{B}_{i+2,2}\left(c_0^{(\ell)},2!c_1^{(\ell)},\cdots,(i+1)!c_i^{(\ell)}\right)}{(i+2)!}\mathcal{B}\left(\mu_x+\mu_y+2l_3+i+1,\mu_z+2l_4+\frac{1}{2}\right)\\
	&\quad\hspace{5cm}\ds\times\left[\ln2+\Psi^{(d)}\left(\mu_z+2l_4+\frac{1}{2}\right)-\Psi^{(d)}\left(a+2l_3+2l_4+\frac{1}{2}+i\right)\right],
	\ea 
\eeq
and
\beq
	c_i^{(\ell)}=\left\{\ba{ll}\ds\frac{(-\ell)_i(\ell+\a_2+\b_2+1)_i}{i!2^i(\a_2+1)_i},&i\le \ell\\0,&i>\ell\ea\right\}. 
\eeq 
\begin{figure}[t]
	\centering
	\includegraphics[width=18cm,height=10cm]{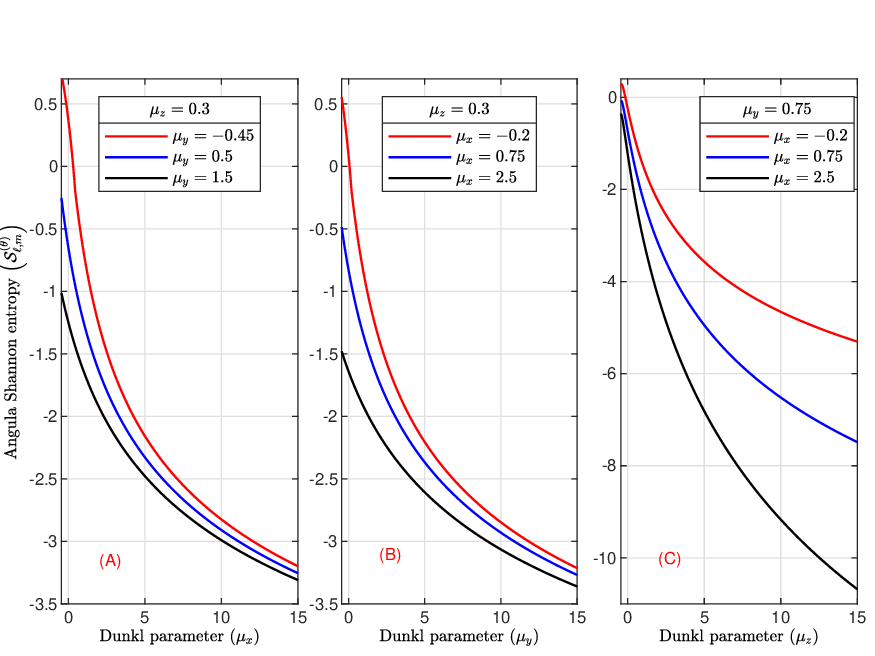}
	\caption{\label{fig4.shannon-th} $\mathcal{S}^{(\theta)}$ of angular density functions $\left[H_{\ell,m}^{(s_3)}(\theta)\right]^2$, of molecules in Dunkl framework. In (A) $\mu_y=-0.45$ (red), 0.5 (blue), 1.5 (black), $\mu_z=0.3$; (B) $\mu_x=-0.2$ (red), 0.75 (blue), 2.5 (black), $\mu_z=0.3$ (C) $\mu_x=-0.2$ (red), 0.75 (blue), 2.5 (black), $\mu_y=0.75$. Other parameters are $\ell=1,m=0$, $s_3=1$.}  
\end{figure}

\subsubsection{Factorization method for $\mathcal{S}_{\ell,m}^{(\theta,1)}$}
	Let $a_{\ell}^{(\a_2,\b_2)}=\frac{(-\ell)_{\ell}(\ell+\a_2+\b_2+1)_{\ell}}{\ell!2^{\ell}(\a_2+1)_{\ell}}$ be the leading term of the Jacobi polynomial, $P_{\ell}^{(\a_2,\b_2)}(q)$ and $-1\le q_1\le q_2\le \cdots\le q_{\ell}\le 1$ denote the roots of the equation $P_{\ell}^{(\a_2,\b_2)}(q)=0$. Then one can write, 
	\beq
	P_{\ell}^{(\a_2,\b_2)}(q)=a_{\ell}^{(\a_2,\b_2)}\prod\limits_{k=1}^{\ell}(q-q_k).
	\eeq
	Using this identity, the integral $\mathcal{S}_{\ell,m}^{(\theta,1)}$, can be written as, 
	\beq
	\ba{ll}
	\mathcal{S}_{\ell,m}^{(\theta,1)}&=-\ds\frac{(N_{\ell,m}^{(\theta)})^2}{2^{a-\frac{1}{2}}}\sum\limits_{j=1}^{\ell}\left[\mathcal{J}_j^{(\theta,R)}+\mathcal{J}_j^{(\theta,L)}\right]-\ln\left[a_{\ell}^{(\a_2,\b_2)}\prod\limits_{j=1}^{\ell}(1-q_j)\right]^2\\
	&-\ds\frac{(N_{\ell,m}^{(\theta)})^2}{2^{a-\frac{1}{2}}}\sum\limits_{j=1}^{\ell}\ln\left(\frac{1+q_j}{1-q_j}\right)^2\frac{2[(\a_2+1)_{\ell}]^2}{[\ell!]^2}\sum\limits_{i=0}^{2\ell}\frac{2^{2l_3+2l_4+i}\widetilde{B}_{i+2,2}\left(c_0^{(\ell)},2!c_1^{(\ell)},\cdots,(i+1)!c_i^{(\ell)}\right)}{(i+2)!}\mathcal{J}_{j,i}^{(\theta)}
	\ea
	\eeq 
	where
	\beq
	\ba{ll}
	\mathcal{J}_j^{(\theta,R)}&=\ds\int_{-1}^{q_j}(1-q)^{\mu_x+\mu_y+2l_3}(1+q)^{\mu_z+2l_4-\frac{1}{2}}\left[P_{\ell}^{(\a_2,\b_2)}(q)\right]^2\ln\left(\frac{q_j-q}{1+q_j}\right)^2dq\\
	\mathcal{J}_j^{(\theta,L)}&=\ds\int_{q_j}^{1}(1-q)^{\mu_x+\mu_y+2l_3}(1+q)^{\mu_z+2l_4-\frac{1}{2}}\left[P_{\ell}^{(\a_2,\b_2)}(q)\right]^2\ln\left(\frac{q-q_j}{1-q_j}\right)^2dq\\
	\mathcal{J}_{j,i}^{(\theta)}&=\ds\int_{-1}^{q_j}(1-q)^{\mu_x+\mu_y+2l_3+i}(1+q)^{\mu_z+2l_4-\frac{1}{2}}dq
	\ea
	\eeq
	By using the transformation $x=\frac{q_j-q}{1+q_j}$, the last integral $\mathcal{J}_{j,i}^{(\theta)}$ can be simplified as, 
	\beq
	\mathcal{J}_{j,i}^{(\theta)}=\ds \sum\limits_{k=0}^{\lfloor\a_2+i\rfloor}\binom{\a_2+i}{k}(1-q_j)^{\a_2+i-k}(1+q_j)^{\b_2+k+1}\mathcal{B}(k+1,\b_2+1).
	\eeq 
	If $(\a_2+i)$ is (+)ve integer, then $\ds\sum\limits_{k=0}^{\lfloor\a_2+i\rfloor}$ is a finite series sum over $k=0,1,2,\cdots,\a_2+i$. Otherwise it is a convergent infinite series. Similarly, one finds that the integrals $\mathcal{J}_j^{(\theta,R)}$, $\mathcal{J}_j^{(\theta,L)}$ exist and using the transformations $x=\frac{q_j-q}{1+q_j}$ and $y=\frac{q-q_j}{1-q_j}$, they can be expressed as:
	\beq
	\ba{lr}\label{j.j.th.r}
	\mathcal{J}_j^{(\theta,R)}&=\ds\frac{4[(\a_2+1)_{\ell}]^2}{[\ell!]^2}\sum\limits_{i=0}^{2\ell}\sum\limits_{k=0}^{\lfloor\a_2+i\rfloor}\left[\frac{2^{2l_3+2l_4+i}\widetilde{B}_{i+2,2}\left(c_0^{(\ell)},2!c_1^{(\ell)},\cdots,(i+1)!c_i^{(\ell)}\right)}{(i+2)!}\binom{\a_2+i}{k}(1-q_j)^{\a_2+i-k}(1+q_j)^{\b_2+k+1}\right]\\
	&\quad\times\mathcal{B}(k+1,\b_2+1)\left[\Psi^{(d)}(k+1)-\Psi^{(d)}(\b_2+k+2)\right],
	\ea
	\eeq 
	and 
	\beq
	\ba{lr}\label{j.j.th.l}
	\mathcal{J}_j^{(\theta,L)}&=\ds\frac{4[(\a_2+1)_{\ell}]^2}{[\ell!]^2}\sum\limits_{i=0}^{2\ell}\sum\limits_{k=0}^{\lfloor\b_2\rfloor}\left[\frac{2^{2l_3+2l_4+i}\widetilde{B}_{i+2,2}\left(c_0^{(\ell)},2!c_1^{(\ell)},\cdots,(i+1)!c_i^{(\ell)}\right)}{(i+2)!}\binom{\b_2}{k}(1-q_j)^{\a_2+i+k+1}(1+q_j)^{\b_2-k}\right]\\
	&\quad\times\mathcal{B}(k+1,\a_2+1)\left[\Psi^{(d)}(k+1)-\Psi^{(d)}(\a_2+k+2)\right].
	\ea
	\eeq
	The series $\sum\limits_{k=0}^{\lfloor\a_2+i\rfloor}$ is finite in Eq.(\ref{j.j.th.r}) if $(\a_2+i) \in\mathbb{N}$ and finite in Eq.~(\ref{j.j.th.l}) if $\b_2\in\mathbb{N}$; otherwise $\sum\limits_{k=0}^{\lfloor\b_2\rfloor}$ they are convergent infinite series.
	
\begin{figure}[t]
	\centering
	\includegraphics[width=18cm,height=10cm]{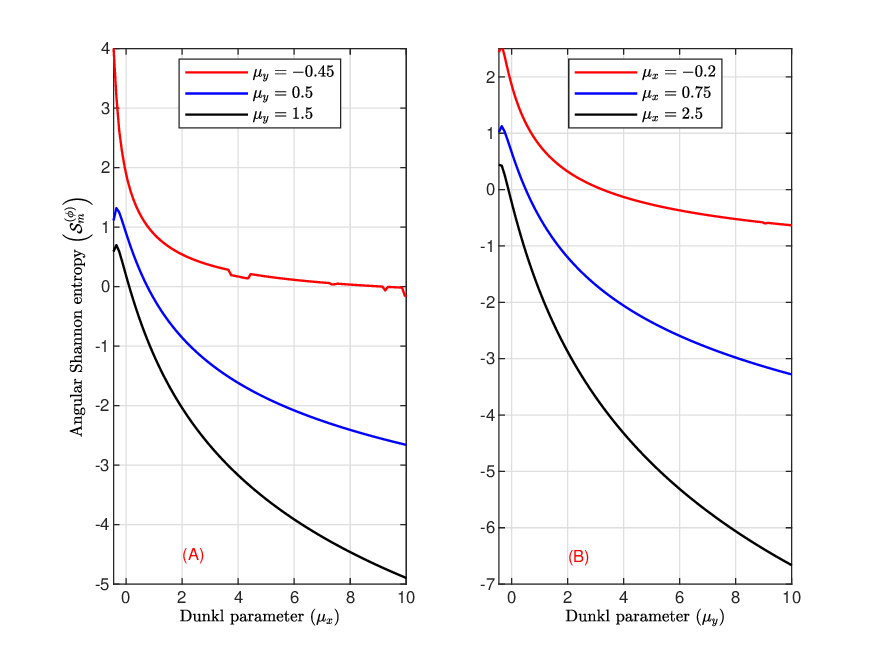}
	\caption{\label{fig5.shannon-phi} Comparison of $S$ of angular density function, $\left[G_{m}^{(s_1,s_2)}(\phi)\right]^2$, for different values of Dunkl parameters. In (A) $\mu_y=-0.45$ (red), 0.5 (blue), 1.5 (black); (B) $\mu_x=-0.2$ (red), 0.75 (blue), 2.5 (black). The other parameters are $m=1$ and $s_1=s_2=1$.}
\end{figure}

\subsection{Shannon entropy of the marginal density function $\left[G_{m}^{(s_1,s_2)}(\phi)\right]^2$}
\noindent Likewise, $\mathcal{S}_m^{(\phi)}$ of the density function $\left[G_{m}^{(s_1,s_2)}(\phi)\right]^2$ can be expressed as,  
	\beq\label{sh.phi.nu}
	\mathcal{S}_m^{(\phi)}=-\ds\int_{0}^{2\pi}\left[G_{m}^{(s_1,s_2)}(\phi)\right]^2\ln[\left[G_{m}^{(s_1,s_2)}(\phi)\right]^2]d\chi_{\phi}.
	\eeq
	This can be rewritten as, 
	\beq\label{sh.phi.qa}
	\mathcal{S}_m^{(\phi)}=\mathcal{S}_m^{(\phi,1)}-\ln[(N_{m}^{(\phi)})^2]+(N_{m}^{(\phi)})^2\left[2l_1\mathcal{S}^{(\phi,2)}+2l_2\mathcal{S}^{(\phi,3)}\right], 
	\eeq
	where
	\beq\label{sh.phi1}
	\ba{ll}
	\mathcal{S}_m^{(\phi,1)}&=\ds-\frac{(N_{m}^{(\phi)})^2}{2^{\mu_x+\mu_y-1}}\ds\int_{-1}^{1}(1-p)^{\mu_y+2l_1-\frac{1}{2}}(1+p)^{\mu_x+2l_2-\frac{1}{2}}\left[P_{m}^{(\a_1,\b_1)}(p)\right]^2\ln\left(\left[P_{m}^{(\a_1,\b_1)}(p)\right]^2\right)dp,
	\ea 
	\eeq
	\beq\label{sh.phi2}
	\ba{ll}
	\mathcal{S}^{(\phi,2)}&=-\ds\frac{1}{2^{\mu_x+\mu_y-1}}\ds\int_{-1}^{1}(1-p)^{\mu_y+2l_1-\frac{1}{2}}(1+p)^{\mu_x+2l_2-\frac{1}{2}}\left[P_{m}^{(\a_1,\b_1)}(p)\right]^2\ln\left[1-p\right]dp\\
	&=-\ds\frac{2[(\a_1+1)_m]^2}{[m!]^2}\sum\limits_{i=0}^{2m}\frac{2^{2l_1+2l_2+i+1}\widetilde{B}_{i+2,2}\left(c_0^{(m)},2!c_1^{(m)},\cdots,(i+1)!c_i^{(m)}\right)}{(i+2)!}\mathcal{B}\left(\mu_y+2l_1+i+\frac{1}{2},\mu_x+2l_2+\frac{1}{2}\right)\\
	&\quad\hspace{5cm}\ds\times\left[\ln2+\Psi^{(d)}\left(\mu_y+2l_1+i+\frac{1}{2}\right)-\Psi^{(d)}\left(\mu_x+\mu_y+2l_1+2l_2+i+1\right)\right],
	\ea
	\eeq
	\beq\label{sh.phi3}
	\ba{ll}
	\mathcal{S}^{(\phi,3)}&=\ds-\frac{1}{2^{\mu_x+\mu_y-1}}\ds\int_{-1}^{1}(1-p)^{\mu_y+2l_1-\frac{1}{2}}(1+p)^{\mu_x+2l_2-\frac{1}{2}}\left[P_{m}^{(\a_1,\b_1)}(p)\right]^2\ln\left[1+p\right]dp\\
	&=-\ds\frac{2[(\a_1+1)_m]^2}{[m!]^2}\sum\limits_{i=0}^{2m}\frac{2^{2l_1+2l_2+i+1}\widetilde{B}_{i+2,2}\left(c_0^{(m)},2!c_1^{(m)},\cdots,(i+1)!c_i^{(m)}\right)}{(i+2)!}\mathcal{B}\left(\mu_y+2l_1+i+\frac{1}{2},\mu_x+2l_2+\frac{1}{2}\right)\\
	&\quad\hspace{5cm}\ds\times\left[\ln2+\Psi^{(d)}\left(\mu_x+2l_2+\frac{1}{2}\right)-\Psi^{(d)}\left(\mu_x+\mu_y+2l_1+2l_2+i+1\right)\right],
	\ea 
	\eeq
	and
	\beq
	c_i^{(m)}=\left\{\ba{ll}\ds\frac{(-m)_i(m+\a_1+\b_1+1)_i}{i!2^i(\a_1+1)_i},&i\le m\\0,&i>m\ea\right\}. 
	\eeq 
	
	\subsubsection{Factorization method for $\mathcal{S}_m^{(\phi,1)}$}
	Let $a_{m}^{(\a_1,\b_1)}=\frac{(-m)_{m}(m+\a_1+\b_1+1)_{m}}{m! \,2^{m}(\a_1+1)_{m}}$ be the leading term of the Jacobi polynomial, $P_{m}^{(\a_1,\b_1)}(p)$, and $-1\le p_1\le p_2\le \cdots\le p_m\le 1$ represent the roots of the equation $P_{m}^{(\a_1,\b_1)}(p)=0$. Then one can write, 
		\beq
		P_{m}^{(\a_1,\b_1)}(q)=a_{m}^{(\a_1,\b_1)}\prod\limits_{k=1}^{\ell}(p-p_k).
		\eeq
		One can then express the integral $\mathcal{S}_{m}^{(\phi,1)}$ as below,  
		\beq
		\ba{ll}
		\mathcal{S}_{m}^{(\phi,1)}&=-\ds\frac{(N_{m}^{(\phi)})^2}{2^{\mu_x+\mu_y-1}}\sum\limits_{j=1}^{m}\left[\mathcal{J}_j^{(\phi,R)}+\mathcal{J}_j^{(\phi,L)}\right]-\ln\left[a_{m}^{(\a_1,\b_1)}\prod\limits_{j=1}^{m}(1-p_j)\right]^2\\
		&-\ds\frac{(N_{m}^{(\phi)})^2}{2^{\mu_x+\mu_y-1}}\sum\limits_{j=1}^{m}\ln\left(\frac{1+p_j}{1-p_j}\right)^2\frac{2[(\a_1+1)_{m}]^2}{[m!]^2}\sum\limits_{i=0}^{2m}\frac{2^{2l_1+2l_2+i}\widetilde{B}_{i+2,2}\left(c_0^{(m)},2!c_1^{(m)},\cdots,(i+1)!c_i^{(m)}\right)}{(i+2)!}\mathcal{J}_{j,i}^{(\phi)},
		\ea
		\eeq 
		where
		\beq
		\ba{ll}
		\mathcal{J}_j^{(\phi,R)}
		&=\ds\frac{2[(\a_1+1)_m]^2}{[m!]^2}\sum\limits_{i=0}^{2m}\sum\limits_{k=0}^{\lfloor\a_1+i\rfloor}\left[\frac{2^{2l_1+2l_2+i+1}\widetilde{B}_{i+2,2}\left(c_0^{(m)},2!c_1^{(m)},\cdots,(i+1)!c_i^{(m)}\right)}{(i+2)!}\binom{\a_1+i}{k}(1-p_j)^{\a_1+i-k}(1+p_j)^{\b_1+k+1}\right]\\
		&\quad\times\mathcal{B}(k+1,\b_1+1)\left[\Psi^{(d)}(k+1)-\Psi^{(d)}(\b_1+k+2)\right],\\
		\mathcal{J}_j^{(\phi,L)}
		&=\ds\frac{2[(\a_1+1)_m]^2}{[m!]^2}\sum\limits_{i=0}^{2m}\sum\limits_{k=0}^{\lfloor\b_1\rfloor}\left[\frac{2^{2l_1+2l_2+i+1}\widetilde{B}_{i+2,2}\left(c_0^{(m)},2!c_1^{(m)},\cdots,(i+1)!c_i^{(m)}\right)}{(i+2)!}\binom{\b_1}{k}(1-p_j)^{\a_1+i+k+1}(1+p_j)^{\b_1-k}\right]\\
		&\quad\times\mathcal{B}(k+1,\a_1+1)\left[\Psi^{(d)}(k+1)-\Psi^{(d)}(\a_1+k+2)\right],\\
		\mathcal{J}_{j,i}^{(\phi)}
		&=\sum\limits_{k=0}^{\lfloor\a_1+i\rfloor}\binom{\a_1+i}{k}(1-p_j)^{\a_1+i-k}(1+p_j)^{\b_1+k+1}\mathcal{B}(k+1,\b_1+1).
		\ea
		\eeq
		Note that here, $\a_i+i\in\mathbb{N}$, $\b_1\in\mathbb{N}$. 

\begin{figure}[t]
	\centering
	\includegraphics[width=18cm,height=10cm]{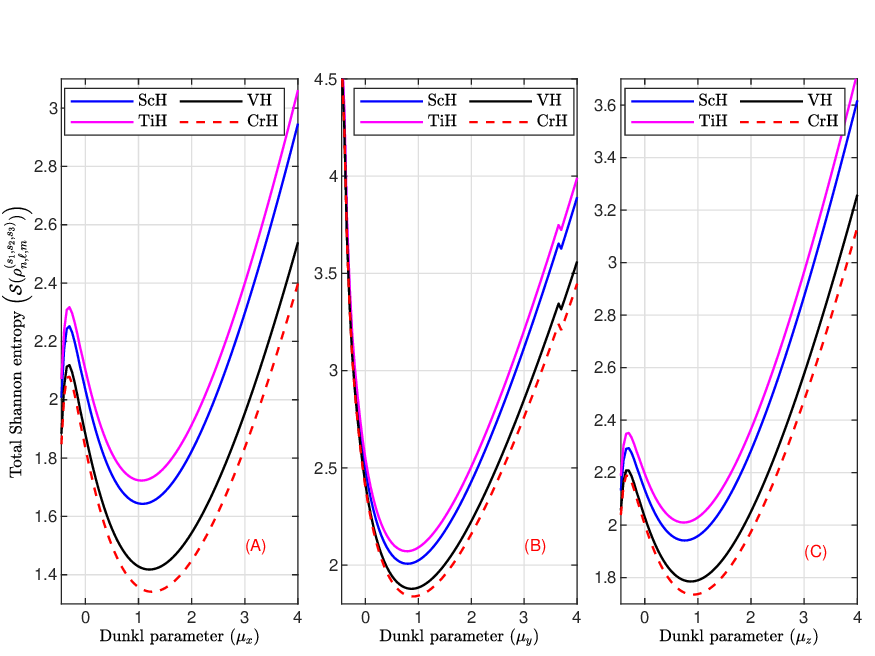}
	\caption{\label{fig6.shannon-r-th-phi} Comparison of $S$ of total density functions $\rho_{n,\ell,m}^{(s_1,s_2,s_3)}$ of four molecules for different values of Dunkl parameters. In (A) $\mu_y=0.75$, $\mu_z=0.3$ (B) $\mu_x=0.5$, $\mu_z=0.3$ and (C) $\mu_x=0.5$, $\mu_y=0.75$. The other parameters are $n=m=\ell=1$, $s_1=s_2=s_3=1$ and $\lambda=0$. } 
\end{figure}

Now let us present some sample results for $S$. For this, the radial and angular components, namely, $\mathcal{S}^{(r)}, \mathcal{S}^{(\theta)}, \mathcal{S}^{(\phi)}$, and composite entropy, $\mathcal{S}$, are reported for TiH molecule in Table~\ref{table3.tih} for $s_1=s_2=s_3=1$. The values of these quantities are obtained from expressions given in Eqs.~(\ref{sha.r.qa}), (\ref{sha.theta.qa}) and (\ref{sh.phi.qa}), while the same for composite Shannon entropy is defined by: $\mathcal{S}=\mathcal{S}^{(r)}+\mathcal{S}^{(\theta)}+\mathcal{S}^{(\phi)}$. It may be noted that, due to the presence of central-field nature of the potential, angular density functions $\left[H_{\ell,m}^{(s_3)}(\theta)\right]^2$ and $\left[G_{m}^{(s_1,s_2)}(\phi)\right]^2$ remain independent of molecular parameters, $D_e,r_e, \alpha$ and $\mu$. But they depend on the Dunkl parameters due to presence of reflection operators. It is worth mentioning that, 
in ground state, all $\lambda$'s, reproduce the corresponding entropies practically exactly. For excited states also, one sees quite 
satisfactory general agreement in entropies for different $\lambda$'s, but slight variations are observed as $\lambda$ is varied. 
This normally holds true for all the molecules considered here. As in previous table, here also, the left, middle and right segments correspond to similar Dunkl parameters. No result could be found out for $S$ in the literature for direct comparison, and we hope that these results would inspire future works in this direction. 

To probe this in detail, Fig.~\ref{fig3.shannon-r} graphically displays the effect of Dunkl parameters on $\mathcal{S}^{(r)}_{n,\ell,m}$, for all four molecules. Representative values are given for $n=\ell=m=1$ state, for a fixed $\lambda =0$ of even parity $s_1=s_2=s_3=1$. The three panels represent changes with respect to $\mu_x$ keeping $\mu_y, \mu_z$ fixed at 0.75 and 0.3 respectively in (A); $\mu_y$ keeping $\mu_x, \mu_z$ fixed at $-0.45$ and 0.3 respectively in (B); and $\mu_z$ for $\mu_x=-0.45$, $\mu_y=0.75$ in (C). This illustrates that $\mathcal{S}^{(r)}_{n,\ell,m}$ is smooth linear increasing function with respect to $\mu_x$, $\mu_y$ and $\mu_z$. Similarly, the effect of Dunkl parameters on $\mathcal{S}_{\ell,m}^{(\theta)}$ of angular density functions, $\left[H_{\ell,m}^{(s_3)}(\theta)\right]^2$ and $\left[G_{m}^{(s_1,s_2)}(\phi)\right]^2$ are produced in Figs.~\ref{fig4.shannon-th} and \ref{fig5.shannon-phi} respectively. As these do not depend on spectroscopic parameters, they remain unchanged from molecule to molecule. From these two figures one can observe that the angular Shannon entropies are monotone decreasing functions with respect to Dunkl parameters. Note that $\mathcal{S}^{(\phi)}_m$ has no dependence of $\mu_z$. Finally, the effect of Dunkl parameters on the total density is shown in Fig. \ref{fig6.shannon-r-th-phi} for all four molecules. From this figure, it is clear that the composite $\mathcal{S}\left(\rho_{n,\ell,m}^{(s_1,s_2,s_3)}\right)$ always passes through some optimum value at certain Dunkl parameter, and after that it gradually increases. 

\begin{figure}[h]
		\centering
		\includegraphics[width=18cm,height=14cm]{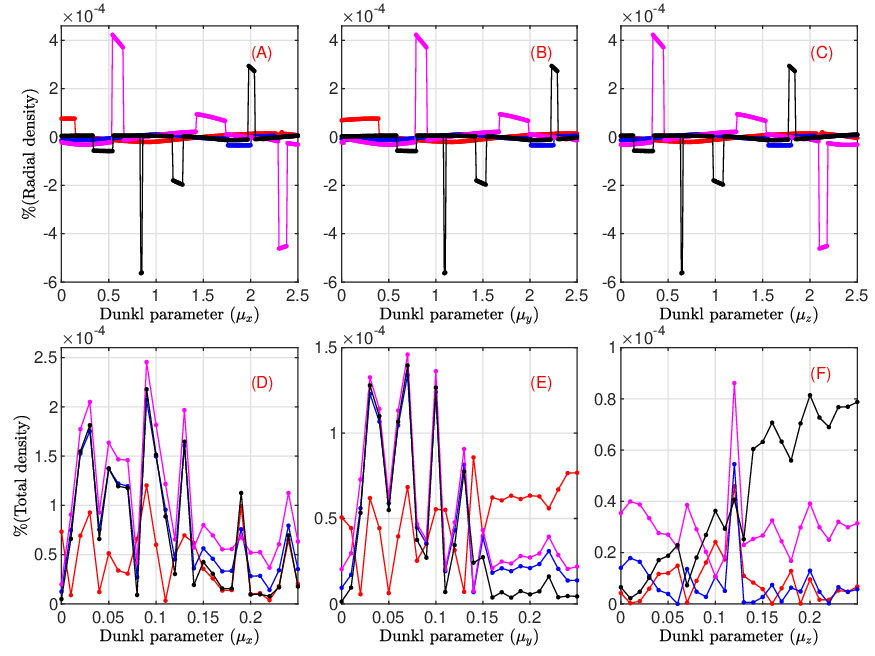}
		\caption{\label{fig7.error-shannon} Comparison of percentage of error of $S$ of radial density function for four molecules with different values of Dunkl parameters.  Left column (A), (D) $\mu_y = 0.75, \mu_z = 0.3$; middle column (B), (E) $\mu_x = 0.5, \mu_z = 0.3$; and right column (C), (F) $\mu_x = 0.5, \mu_y = 0.75$. The red, blue, black, magenta lines represent percent error of $S$ of ScH, TiH, VH and CrH respectively. The quantum numbers are $n=\ell=m=1$ and$\lambda=0$. } 
	\end{figure}

\begin{figure}[h]
		\centering
		\includegraphics[width=18cm,height=14cm]{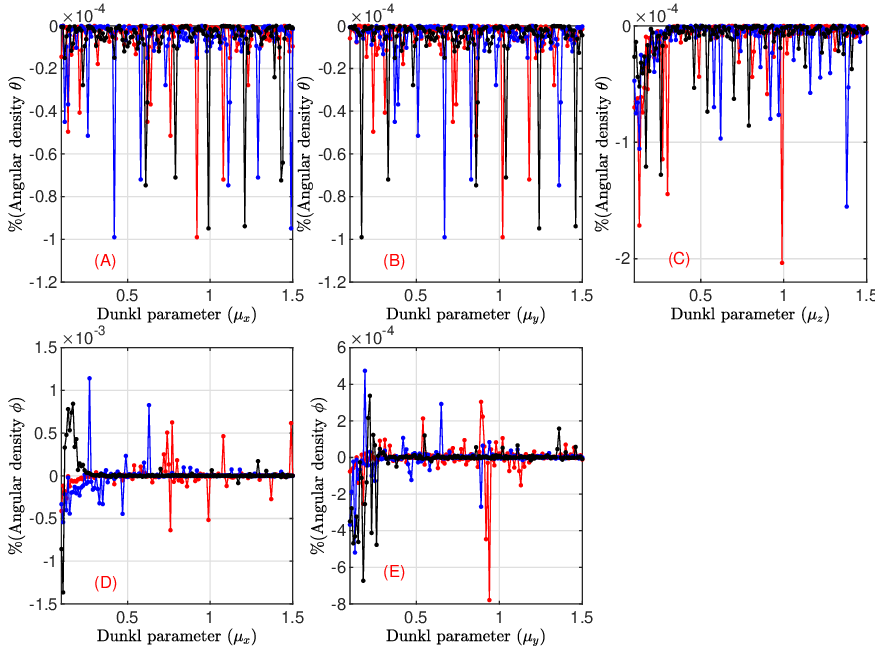}
		\caption{\label{fig8.error-thphi} Comparison of percent error of $S$ of angular density functions for different values of Dunkl parameters. In (A), (D) percent error of $S$ of angular density functions with respect to Dunkl parameter $\mu_x$ where red, blue, black lines are drawn for $\mu_y=0.5, 1, 1.5$ respectively keeping $\mu_z=0.6$ fixed. Similarly, in (B), (E) with respect to $\mu_y$ for fixed $\mu_z=0.6$. In (C) with respect to $\mu_z$ for fixed $\mu_y=0.75$, where red, blue, black lines are drawn for $\mu_x=0.4, 0.75, 1.25$ respectively. The quantum numbers are $\ell=1$, $m=1$ and $\lambda=0$.} 
	\end{figure}

Before passing, a few remarks may be made regarding the calculation of the entropies. We note that, some of the 
integrals in Eqs.~(\ref{sha.r1}), (\ref{sha.theta1}), and (\ref{sh.phi1}) can be expressed directly into analytic forms, 
by means of the factorization method involving sum of convergent infinite series. As an additional check, we have also 
performed the integration in Eqs.~(\ref{sha.r.nu}), (\ref{sha.theta.nu}) and (\ref{sh.phi.nu}) numerically, resulting in
the so-called semi-analytical estimates. The per cent $\epsilon_{\%} (\rho)$ error of numerical ($\mathcal{S}_{\textrm{num}}$), 
with respect to semi-analytical ($\mathcal{S}(\rho)$) of a density function can be estimated from: 
$\epsilon_{\%}(\rho)=\frac{|\mathcal{S}(\rho)-\mathcal{S}_{\textrm{num}}(\rho)|}{\mathcal{S}_(\rho)}\times100$. A detailed 
analysis has been made for this on all states of all molecules; it reveals that the two results are indeed very close. To 
cite an example, for TiH molecule, one finds, $\epsilon_{\%}\left([\frac{R_{1,1,1}(r)}{r^{a}}]^2\right)=0.00004$, 
$\epsilon_{\%}([H_{1,1}^{(s_3)}]^2)=0.000009$, $\epsilon_{\%}([G_{1}^{(s_1,s_2)}]^2)=0.0007$ for 
$\mu_x=0.03, \mu_y=0.01,\mu_z=0.03$ and $\lambda=-0.25$. The percentage error of total density function is $\epsilon_{\%}(\rho_{1,1,1}^{(s_1,s_2,s_3)})=0.00004$. One can observe that $S$ of total density function is the sum of marginal density functions but $\epsilon_{\%}(\rho_{1,1,1}^{(s_1,s_2,s_3)})\ne \epsilon_{\%}\left([\frac{R_{1,1,1}(r)}{r^{a}}]^2\right)+\epsilon_{\%}([H_{1,1}^{(s_3)}]^2)+\epsilon_{\%}([G_{1}^{(s_1,s_2)}]^2)$. The numerical values of $\epsilon_{\%}(\rho_{n,\ell,m}^{(s_1,s_2,s_3)})$ error of Shannon entropy of joint density function of TiH molecule are shown in table \ref{table3.tih} for three sets of Dunkl parameters. In Fig. \ref{fig7.error-shannon} (A), (B), (C) we have plotted the percentage error of $S$ of total density function with respect to the Dunkl parameter $\mu_x$, $\mu_y$ and $\mu_z$ respectively. On the other-hand percentage error of $S$ of radial density function are plotted with respect to the same in Fig.~\ref{fig7.error-shannon} (D), (E), (F). Similarly, percentage error of $S$ of angular density function $\theta$ are plotted in \ref{fig8.error-thphi} (A), (B), (C) and of $\phi$ plotted in Fig. \ref{fig8.error-thphi} (D), (E). From these two figures we observe that the absolute per cent deviation between the analytical and numerical results remain well within 0.0001\%.

\begin{figure}[h]
	\centering
	\includegraphics[width=18cm,height=10cm]{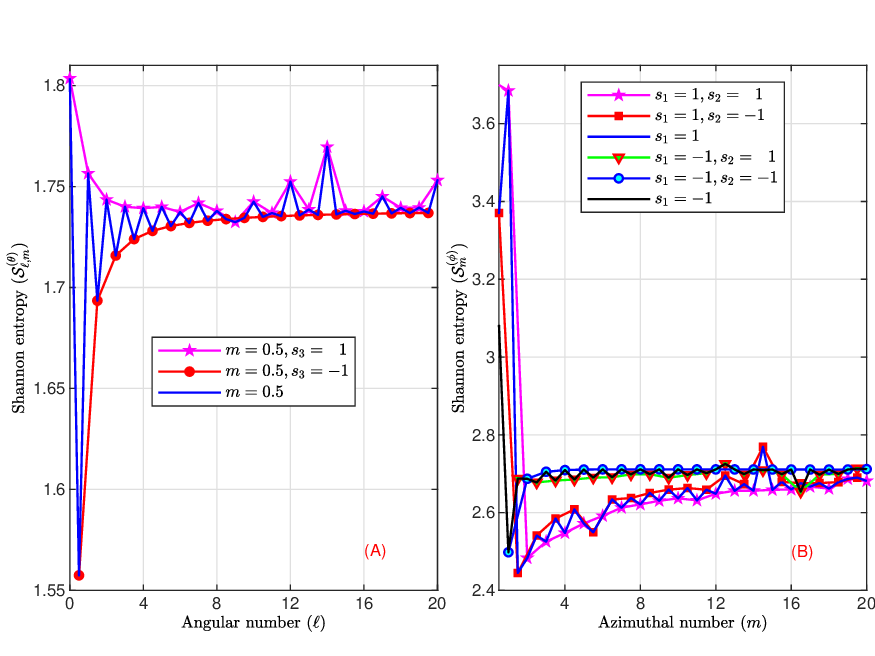}
	\caption{\label{fig10.reflection.shannon} Comparison of the effect of reflection operator on $\mathcal{S}_{\ell,m}^{(\theta)}$ and $\mathcal{S}_m^{(\phi)}$ with respect to the angular quantum number $\ell$ in (A) and azimuthal number $m$ in (B). Common parameters are $\mu_x=-0.45,\mu_y=-0.4,\mu_z=-0.3$.} 
	\end{figure}

In absence of $\widehat{R}_x, \widehat{R}_y, \widehat{R}_z$ and $\mu_x, \mu_y, \mu_z$, quasi analytical value of $S$ of Legendre \cite{dehesa.fisher.sannon2005,dehesa.spherical2007}, hypergeometric \cite{dehesa.hypergeometric2013}, hyperspherical \cite{log.pot,dehesa.spherical2007}, Jacobi \cite{dehesa.jacobi2010,dehesa.hypergeometric2013}, Laguerre \cite{dehesa.fisher.sannon2005,dehesa.ijqc2011,dehesa.laguerre2011,dehesa.hypergeometric2013} polynomials have been defined in the literature; for these and other problems, we refer to the reader to \cite{log.pot,shannon2}. In this manuscript, we have obtained analytical value of Shannon entropy in Schr\"odinger-Dunkl framework, which is different from the usual Schr\"odinger system. Due to presence of $\widehat{R}_x$ in $N_{\theta}$ and $\widehat{R}_y, \widehat{R}_z$ in $B_{\phi}$, the angular wave solutions $H_{\ell,m}^{(s_3)}(\theta)$ and $G_m^{(s_1,s_2)}(\phi)$ are directly impacted by them. The radial wave solution depends only on the parity values $s_1,s_2,s_3=\pm1$, but not directly affected by $\widehat{R}_x, \widehat{R}_y, \widehat{R}_z$. To realize the effect of $\widehat{R}_x, \widehat{R}_y, \widehat{R}_z$ on Shannon entropy of the angular wave solutions we have plotted Shannon entropies with respect to $\ell$ and $m$ quantum numbers in Fig.~\ref{fig10.reflection.shannon} for $\mu_x=-0.45$, $\mu_y=-0.4$, $\mu_z=-0.3$. In Fig. \ref{fig10.reflection.shannon} (A) if $s_3=1$, then $\ell\in \mathbb{N}\cup\{0\}$; if $s_3=-1$, then $\ell\in\frac{1}{2}+\mathbb{N}\cup\{0\}$. In this case $H_{\ell,m}^{(s_3)}(\theta)=N_{\ell0l_4,m}^{(\theta)}\cos^{e_3}\theta\sin^{2m}\theta\,P_{\ell-l_4}^{(\a_2,\b_2)}(\cos2\theta)$. In Fig. \ref{fig10.reflection.shannon} (B), if $s_1s_2=-1$, $m\in\frac{1}{2}+\mathbb{N}\cup\{0\}$; if $s_1s_2=1$, $m\in\mathbb{N}$ and $m=0$ only when $s_1=s_2=1$. In this case $G_m^{(s_1,s_2)}(\phi)=N_{m-l_1-l_2}^{(\phi)}\sin^{2l_1}\phi\cos^{2l_2}\phi\,P_{m-l_1-l_2}^{(\a_1,\b_1)}(\cos2\phi)$. From this figure it is clear that the Shannon entropy oscillates due to reflections. But one can observe that, without $\mu_x, \mu_y, \mu_z$ and $\widehat{R}_x, \widehat{R}_y, \widehat{R}_z$, $\mathcal{S}_{\ell,m}^{(\theta)}$ increases when $\ell$ grows, but $\mathcal{S}_m^{(\phi)}$ has a fixed value for all $m$.  

	
\begin{figure}[t]
	\centering
	\includegraphics[width=18cm,height=10cm]{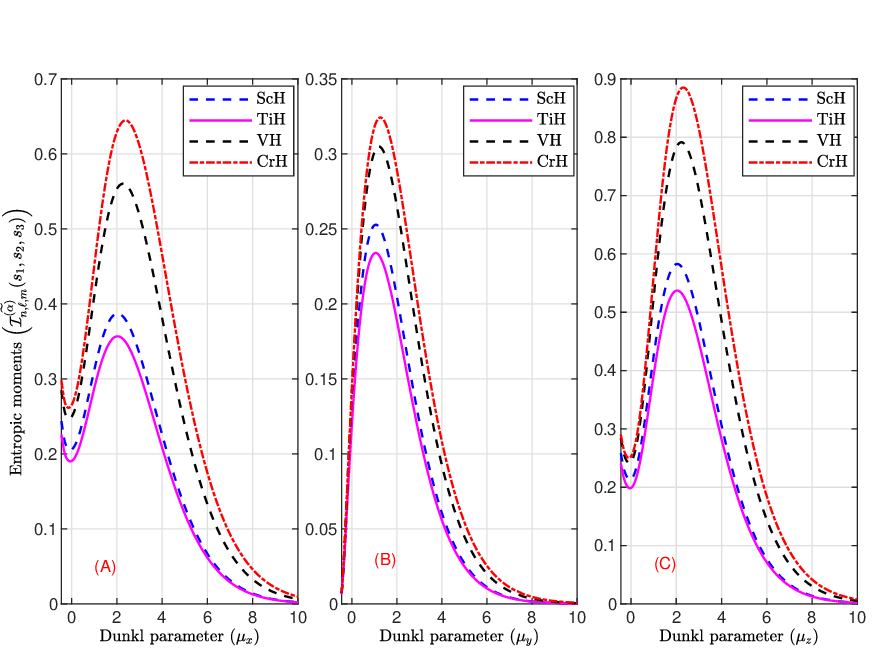}
	\caption{\label{fig9.moments} Comparison of entropic moment of order $\widetilde{\a}=2.25$ with respect to Dunkl parameters, in $n=\ell=m=1$ state. In (A) $\mu_y=0.75,\mu_z=0.3$, (B) $\mu_x=-0.45,\mu_z=0.3$ and (C) $\mu_x=-0.45,\mu_y=0.75$. Common parameters are: $s_1=s_2=s_3=1$, $\lambda=-0.25$. }
\end{figure}

\section{Moment generating functions and information}\label{sec6.moment}
\subsection{Entropic moment, Disequilibrium}
The entropic moment of the density function $\rho_{n,\ell,m}^{(s_1,s_2,s_3)} (r, \theta, \phi)$ for an arbitrary state $\psi_{n,\ell,m}^{(s_1,s_2,s_3)}(r,\theta,\phi)$ is given as, 
\beq
\mathcal{I}^{(\widetilde{\a})}_{n,\ell,m}(s_1,s_2,s_3)=\ds\int\left[\rho_{n,\ell,m}^{(s_1,s_2,s_3)}\right]^{\widetilde{\a}}d\chi=\ds\mathcal{I}^{(r,\widetilde{\a})}_{n,\ell,m}\mathcal{I}^{(\theta,\widetilde{\a})}_{\ell,m}(s_3)\mathcal{I}^{(\phi,\widetilde{\a})}_{m}(s_1,s_2)
\eeq 
where $\mathcal{I}^{(r,\widetilde{\a})}_{n,\ell,m}$, $\mathcal{I}^{(\theta,\widetilde{\a})}_{\ell,m}(s_3)$ and $\mathcal{I}^{(\phi,\widetilde{\a})}_m(s_1,s_2)$ are entropic moments of the marginal density functions. The radial entropic moment of the marginal density function is defined by, 
\beq
\ba{ll}
\mathcal{I}^{(r,\widetilde{\a})}_{n,\ell,m}&=\ds\int_0^{\infty}\left[\frac{R_{n,\ell,m}(r)}{r^{a}}\right]^{2\widetilde{\a}}d\chi_{r}\\
&=\ds-\frac{[N^{(r)}_{n,\ell,m}]^{2\widetilde{\a}}}{(-\a)^{2a(\widetilde{\a}-1)+1}}\sum\limits_{i=0}^{2n\widetilde{\a}}\frac{(2\widetilde{\a})!\widetilde{B}_{2\widetilde{\a}+i,2\widetilde{\a}}\left(c_0^{(n)},2!c_1^{(n)},\cdots,(i+1)!c_i^{(n)}\right)}{(2\widetilde{\a}+i)!}\int_0^1s^{2\widetilde{\a}\varepsilon+i-1}(1-s)^{2L\widetilde{\a}}\left[\ln(s)\right]^{2a(\widetilde{\a}-1)}ds.~
\ea 
\eeq
In a particular case, if $2a(\widetilde{\a}-1)\in\mathbb{N}$, then we can find the analytical form of $\mathcal{I}^{(r,\widetilde{\a})}_{n,\ell,m}$; otherwise it needs to be treated numerically. The analytical form is given in differentiation form as: 
\beq
\mathcal{I}^{(r,\widetilde{\a})}_{n,\ell,m}=\ds-\frac{[N^{(r)}_{n,\ell,m}]^{2\widetilde{\a}}}{(-\a)^{2a(\widetilde{\a}-1)+1}}\sum\limits_{i=0}^{2n\widetilde{\a}}\frac{(2\widetilde{\a})!\widetilde{B}_{2\widetilde{\a}+i,2\widetilde{\a}}\left(c_0^{(n)},2!c_1^{(n)},\cdots,(i+1)!c_i^{(n)}\right)}{(2\widetilde{\a}+i)!}\lim\limits_{p\rightarrow0}\frac{d^{2a(\widetilde{\a}-1)}}{dp^{2a(\widetilde{\a}-1)}}\mathcal{B}(2\widetilde{\a}\varepsilon+i+p,2L\widetilde{\a}+1).
\eeq 
The angular entropic moment of the marginal density function $\left[H_{\ell,m}^{(s_3)}(\theta)\right]^2$ can be expressed as, 
\beq
\ba{ll}
\mathcal{I}^{(\theta,\widetilde{\a})}_{\ell,m}(s_3)&=\ds\int_0^{\pi}\left[H_{\ell,m}^{(s_3)}(\theta)\right]^{2\widetilde{\a}}d\chi_{\theta}\\
&=\ds\left[\frac{(\a_2+1)_{\ell}N^{(\theta)}_{n,\ell,m}}{\ell}\right]^{2\widetilde{\a}}\sum\limits_{i=0}^{2\ell\widetilde{\a}}\frac{(2\widetilde{\a})!\widetilde{B}_{2\widetilde{\a}+i,2\widetilde{\a}}\left(c_0^{(\ell)},2!c_1^{(\ell)},\cdots,(i+1)!c_i^{(\ell)}\right)}{(2\widetilde{\a}+i)!}2^{2\widetilde{\a}(l_3+l_4)+i-1}\\
&\quad\hspace{5cm}\ds\times \mathcal{B}\left(\mu_x+\mu_y+2\widetilde{\a} l_3+1+i,\mu_z+2\widetilde{\a} l_4+\frac{1}{2}\right).
\ea
\eeq
Similarly, the angular entropic moment of the marginal density function $\left[G_{m}^{(s_1,s_2)}(\phi)\right]^2$ can be written in the following form, 
\beq
\ba{ll}
\mathcal{I}^{(\phi,\widetilde{\a})}_m(s_1,s_2)&=\ds\int_0^{\pi}\left[G_{m}^{(s_1,s_2)}(\phi)\right]^{2\widetilde{\a}}d\chi_{\phi}\\
&=\ds\left[\frac{(\a_1+1)_mN^{(\phi)}_{n,\ell,m}}{m}\right]^{2\widetilde{\a}}\sum\limits_{i=0}^{2m\widetilde{\a}}\frac{(2\widetilde{\a})!\widetilde{B}_{2\widetilde{\a}+i,2\widetilde{\a}}\left(c_0^{(m)},2!c_1^{(m)},\cdots,(i+1)!c_i^{(m)}\right)2^{2\widetilde{\a}(l_1+l_2) +i+1}}{(2\widetilde{\a}+i)!}\\
&\quad\hspace{5cm}\ds\times \mathcal{B}\left(\mu_y+2\widetilde{\a} l_1+\frac{1}{2}+i,\mu_x+2\widetilde{\a} l_2+\frac{1}{2}\right).
\ea
\eeq
For the particular value of $\widetilde{\a}=2$, the entropic moment reduces to the disequilibrium $\mathcal{D}^{(r,\theta,\phi)}_{n,\ell,m}=\mathcal{I}^{(2)}_{n,\ell,m}$.
The disequilibrium of the joint density function can be written as a product of disequilibrium of marginal density functions, i.e.,  $\mathcal{D}^{(r,\theta,\phi)}_{n,\ell,m}=\mathcal{D}^{(r)}_{n,\ell,m} \mathcal{D}^{(\theta)}_{n,\ell,m} \mathcal{D}^{(\phi)}_{n,\ell,m}$. In Fig.~\ref{fig9.moments}, the entropic moment, $\mathcal{I}^{(\widetilde{\a})}_{n,\ell,m}(s_1,s_2,s_3)$, of the total density function with respect to the Dunkl parameters $\mu_x$, $\mu_y$ and $\mu_z$ are depicted in panels (A), (B) and (C) respectively. Once again the same our hydrides of Sc, Ti, V and Cr are considered. All the molecules show similar trend in all the panels. In terms of qualitative bahavior, the left and right panels show some similarity between the two; namely a minimum followed by a maximum and finally concluded by a gradual decay. On the other hand, the middle panel shows only a maximum followed by a steady decline for the large Dunkl parameter. 

Note that, in absence of $\mu_x, \mu_y, \mu_z$, the angular functions, $H_{\ell,m}^{(s_3)}(\theta)$ and $G_m^{(s_1,s_2)}(\phi)$ reduce to the associated Legendre polynomial and $\frac{e^{im\pi}}{\sqrt{2\pi}}$ respectively. Then one can compare the entopic moments of angular wave functions \cite{dehesa.fisher.sannon2005,dehesa.spherical2007}.
\begin{figure}[t]
	\centering
	\includegraphics[width=18cm,height=10cm]{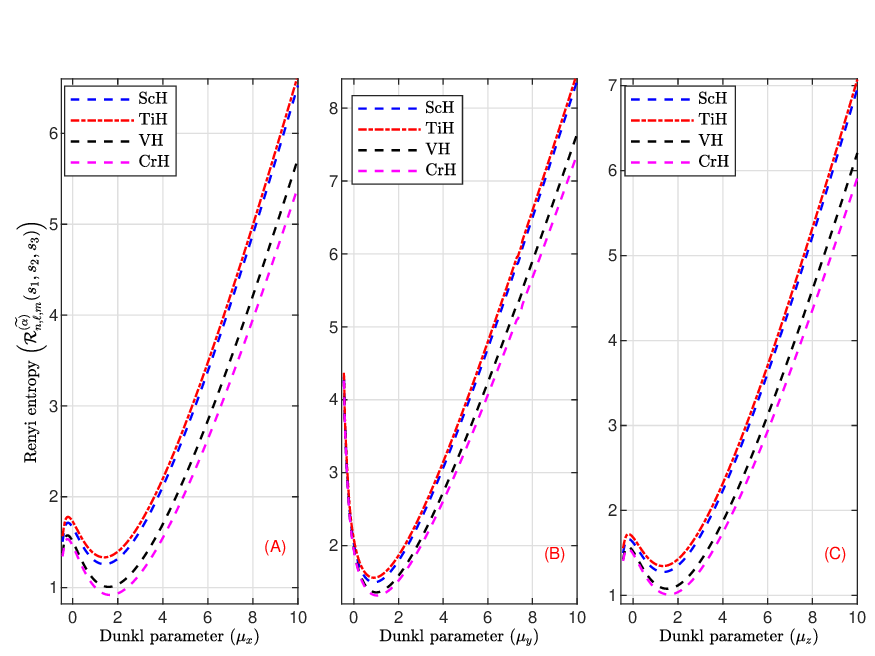}
	\caption{\label{fig10.renyi} Comparison of Renyi entropy of order $\widetilde{\a}= 1.5$ with respect to Dunkl parameters, in $n=\ell=m=1$ state. In (A) $\mu_y=0.75,\mu_z=0.3$, (B) $\mu_x=-0.45,\mu_z=0.3$ and (C) $\mu_x=-0.45,\mu_y=0.75$. Common parameters are: $s_1=s_2=s_3=1$, $\lambda=-0.25$.}
\end{figure}
\begin{figure}[h]
	\centering
	\includegraphics[width=18cm,height=14cm]{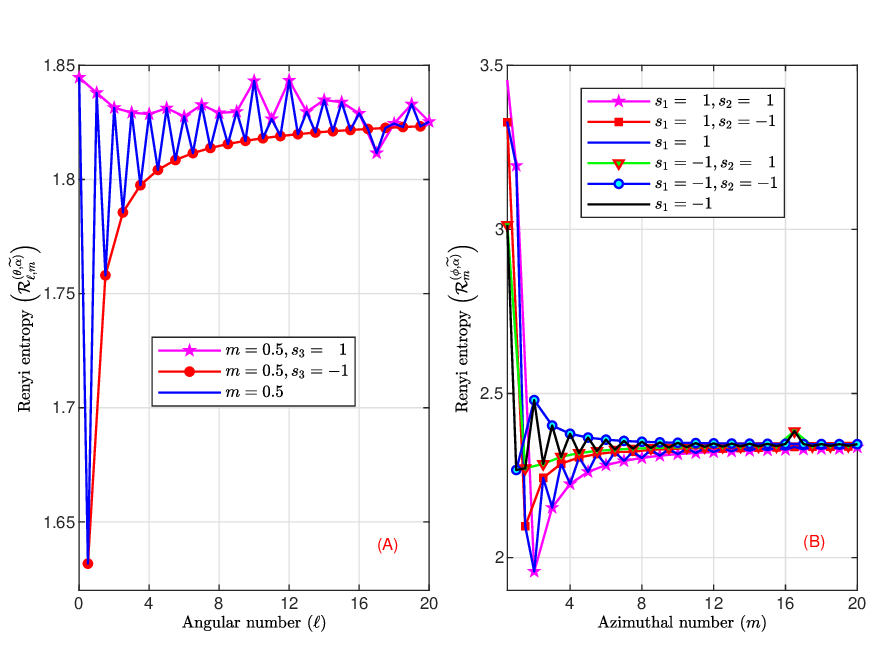}
	\caption{\label{fig13.reflection.renyi} Comparison of the effect of reflection operator on R\'enyi entropies $\mathcal{R}^{(\theta,\widetilde{\a})}_{\ell,m}$ and $\mathcal{R}^{(\phi,\widetilde{\a})}_{m}$ with respect to the angular quantum number $\ell$ in (A) of order $\widetilde{\a}=0.75$ and $m$ in (B) of order $\widetilde{\a}=2.5$. Common parameters are $\mu_x=-0.45,\mu_y=-0.4,\mu_z=-0.3$.} 
\end{figure}

\subsection{R\'enyi entropy}
Now, the R\'enyi entropy of the density function $\rho_{n,\ell,m}^{(s_1,s_2,s_3)}$ is given by \cite{renyi}, 
\beq
\mathcal{R}^{(\widetilde{\a})}_{n,\ell,m}(s_1,s_2,s_3)=\frac{1}{1-\widetilde{\a}}\ln\left[\mathcal{I}^{(\widetilde{\a})}_{n,\ell,m}(s_1,s_2,s_3)\right],\widetilde{\a}>0,\ne 1.
\eeq 
Moreover, the R\'enyi entropy of the joint density function $\rho_{n,\ell,m}^{(s_1,s_2,s_3)}$ can be expressed as a sum of R\'enyi entropies of marginal density functions, as found below,  
\beq
\mathcal{R}^{(\widetilde{\a})}_{n,\ell,m}(s_1,s_2,s_3)=\mathcal{R}^{(r,\widetilde{\a})}_{n,\ell,m}+\mathcal{R}^{(\theta,\widetilde{\a})}_{\ell,m}(s_3)+\mathcal{R}^{(\phi,\widetilde{\a})}_{m}(s_1,s_2)
\eeq 
where 
\beq
\left.
\ba{ll}\mathcal{R}^{(r,\widetilde{\a})}_{n,\ell,m}=\frac{1}{1-\widetilde{\a}}\ln\left[\mathcal{I}^{(r,\widetilde{\a})}_{n,\ell,m}\right],\\
\mathcal{R}^{(\theta,\widetilde{\a})}_{\ell,m}(s_3)=\frac{1}{1-\widetilde{\a}}\ln\left[\mathcal{I}^{(\theta,\widetilde{\a})}_{\ell,m}(s_3)\right]\\
\mathcal{R}^{(\phi,\widetilde{\a})}_{m}(s_1,s_2)=\frac{1}{1-\widetilde{\a}}\ln\left[\mathcal{I}^{(\phi,\widetilde{\a})}_{m}(s_1,s_2)\right]
\ea
\right\} \widetilde{\a}>0,\ne 1.
\eeq
The entropic moments of the marginal density functions are defined analytically for positive integral order. Therefore, R\'enyi entropy of joint density and marginal density functions are obtained analytically for $\widetilde{\a}\in\mathbb{N}-\{1\}$. If $\widetilde{\a}=2$, then R\'enyi entropy $\mathcal{R}^{(\widetilde{\a})}_{n,\ell,m}(s_1,s_2,s_3)$ is connected to the disequilibrium as  $\mathcal{R}^{(\widetilde{\a})}_{n,\ell,m}(s_1,s_2,s_3)=-\ln[\mathcal{D}^{(r,\theta,\phi)}_{n,\ell,m}]$. Disequilibrium is called self similarity \cite{carbo} and it is used in linear entropy \cite{linear}. Here, we have calculated R\'enyi entropy numerically for $\widetilde{\a}=1.5$ and illustrative R\'enyi entropies, $\mathcal{R}^{(\widetilde{\a})}_{n,\ell,m}(s_1,s_2,s_3)$ are graphically shown in Fig.~\ref{fig10.renyi}, for ScH, TiH, VH and CrH molecules. As in the previous figure, here again we see some similarity between the two plots in left and right panels. But in all the panels a sharp minimum is observed for all the molecules. 
If $\mu_x=\mu_y=\mu_z=0$, then for Renyi entropy we refer to the reader in \cite{dehesa.jacobi2010,dehesa.ijqc2011,dehesa.laguerre2011,dehesa.hypergeometric2013}.

To realize the effect of $\widehat{R}_x, \widehat{R}_y, \widehat{R}_z$ on R\'enyi entropy of $H_{\ell,m}^{(s_3)}(\phi)$ and $G_m^{(s_1,s_2)}(\phi)$ we have plotted $\mathcal{R}^{(\theta,0.75)}_{\ell,m}$ and $\mathcal{R}^{(\phi,2.5)}_{m}$ with respect to the angular quantum number $\ell$ in (A) and azimuthal number $m$ in (B) for $\mu_x=-0.45$, $\mu_y=-0.4$, $\mu_z=-0.3$. The R\'enyi entropy oscillates due to reflection but without reflection it is monotone, whose nature depend on $\widetilde{\a}$.   

\begin{figure}[t]
	\centering
	\includegraphics[width=18cm,height=10cm]{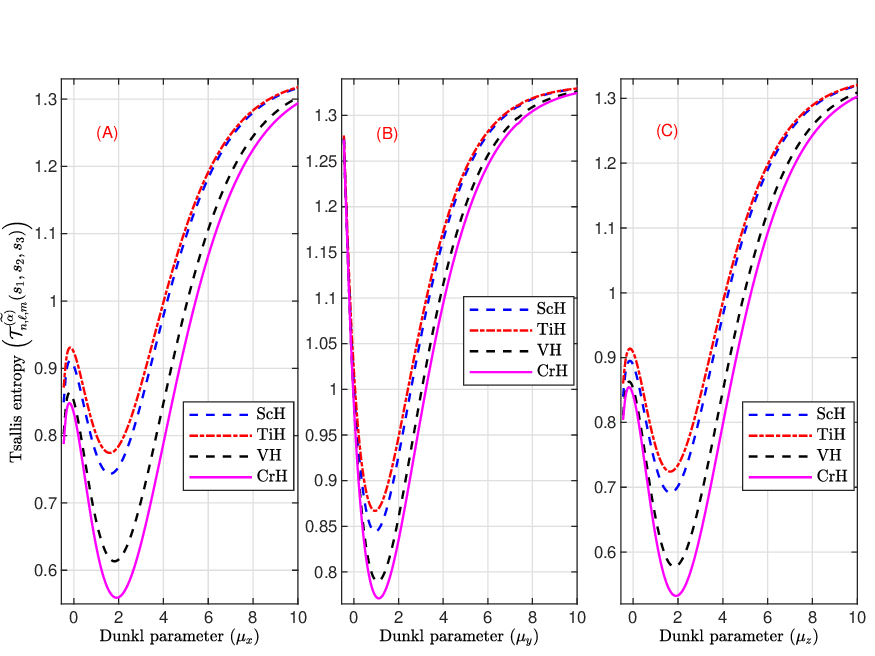}
	\caption{\label{fig11.tsallis} Comparison of Tsallis entropy of order $\widetilde{\a}=1.75$ with respect to Dunkl parameters, in $n=\ell=m=1$ state. In (A) $\mu_y=0.75,\mu_z=0.3$, (B) $\mu_x=-0.45,\mu_z=0.3$ and (C) $\mu_x=-0.45,\mu_y=0.75$. Common parameters are: $s_1=s_2=s_3=1$, $\lambda=0$.}
\end{figure}
\begin{figure}[h]
	\centering
	\includegraphics[width=18cm,height=14cm]{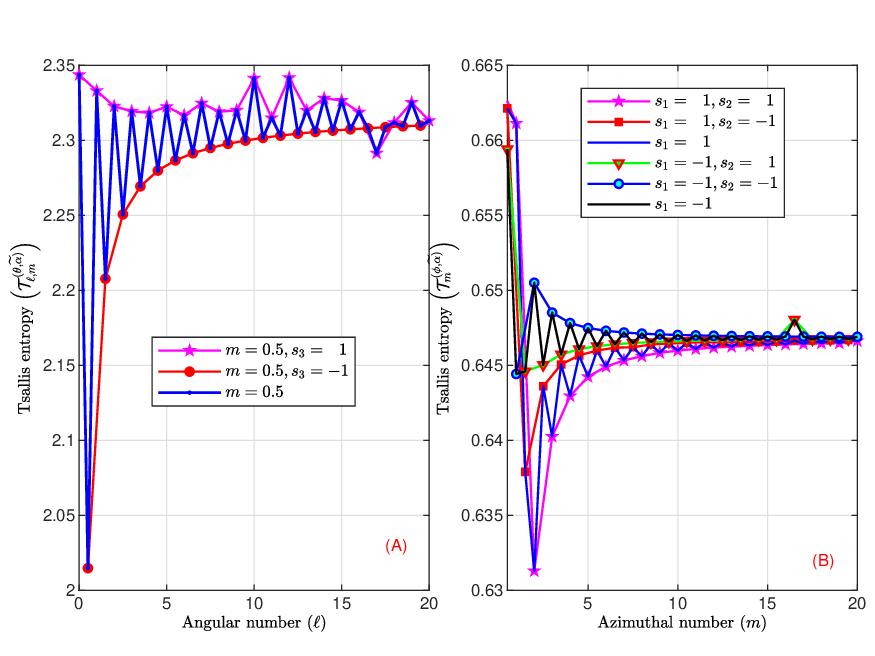}
	\caption{\label{fig15.reflection.tsallis} Comparison of the effect of reflection operator on Tsallis entropies $\mathcal{T}^{(\theta,\widetilde{\a})}_{\ell,m}$ and $\mathcal{T}^{(\phi,\widetilde{\a})}_{m}$ of order $\widetilde{\a}=2.5$ with respect to the angular quantum number $\ell$ in (A) of order $\widetilde{\a}=0.75$ and azimuthal number $m$ in (B) of order $\widetilde{\a}=2.5$. Common parameters are $\mu_x=-0.45,\mu_y=-0.4,\mu_z=-0.3$.} 
\end{figure}

\subsection{Tsallis entropy}  
Lastly, Tsallis entropy of the density function $\rho_{n,\ell,m}$ can be written as \cite{tsallis},
\beq
\mathcal{T}^{(\widetilde{\a})}_{n,\ell,m}(s_1,s_2,s_3)=-\ds\int\left[\rho_{n,\ell,m}^{(s_1,s_2,s_3)}\right]^{\widetilde{\a}}\ln_{\widetilde{\a}}\left[\rho_{n,\ell,m}^{(s_1,s_2,s_3)}\right]d\chi,\widetilde{\a}>0,\ne 1,
\eeq 
where $\ln_{\widetilde{\a}}$ is the $\widetilde{\a}$-logarithm function, defined as $\ln_{\widetilde{\a}}(x)=\frac{x^{1-\widetilde{\a}}-1}{1-\widetilde{\a}}$, for $x>0$, $\widetilde{\a}\in\mathbb{R}-\{1\}$. When $\widetilde{\a}\rightarrow1$, $\ln_{\widetilde{\a}}(x)$ reduces to the natural logarithm function $\ln(x)$.
Then, Tsallis entropy of joint density function $\rho_{n,\ell,m}$ can be expressed as:
\beq
\ba{lr}
\mathcal{T}^{(\widetilde{\a})}_{n,\ell,m}&=\mathcal{T}^{(r,\widetilde{\a})}_{n,\ell,m}+\mathcal{T}^{(\theta,\widetilde{\a})}_{\ell,m}+\mathcal{T}^{(\phi,\widetilde{\a})}_{m}+(1-\widetilde{\a})\left[\mathcal{T}^{(r,\widetilde{\a})}_{n,\ell,m}\mathcal{T}^{(\theta,\widetilde{\a})}_{\ell,m}+\mathcal{T}^{(r,\widetilde{\a})}_{n,\ell,m}\mathcal{T}^{(\phi,\widetilde{\a})}_{m}+\mathcal{T}^{(\theta,\widetilde{\a})}_{\ell,m}\mathcal{T}^{(\phi,\widetilde{\a})}_{m}\right] +(1-\widetilde{\a})^2\mathcal{T}^{(r,\widetilde{\a})}_{n,\ell,m}\mathcal{T}^{(\theta,\widetilde{\a})}_{\ell,m}\mathcal{T}^{(\phi,\widetilde{\a})}_{m}, 
\ea 
\eeq 
where 
\beq
\left.
\ba{ll}
\mathcal{T}^{(r,\widetilde{\a})}_{n,\ell,m}=-\ds\int_0^{\infty}\left[\frac{R_{n,\ell,m}(r)}{r^{a}}\right]^{2\widetilde{\a}}\ln_{\widetilde{\a}}\left[\frac{R_{n,\ell,m}(r)}{r^{a}}\right]^2 d\chi_{r},\\
\mathcal{T}^{(\theta,\widetilde{\a})}_{\ell,m}(s_3)=-\ds\int_0^{\pi}\left[H_{\ell,m}^{(s_3)}(\theta)\right]^{2\widetilde{\a}}\ln_{\widetilde{\a}}\left[H_{\ell,m}^{(s_3)}(\theta)\right]^2 d\chi_{\theta}\\
\mathcal{T}^{(\phi,\widetilde{\a})}_{m}(s_1,s_2)=-\ds\int_0^{2\pi}\left[G_{m}^{(s_1,s_2)}(\phi)\right]^{2\widetilde{\a}}\ln_{\widetilde{\a}}\left[G_{m}^{(s_1,s_2)}(\phi)\right]^{2} d\chi_{\phi}
\ea
\right\} \widetilde{\a}>0,\ne 1,
\eeq
are Tsallis entropies of the marginal density functions $\left[\frac{R_{n,\ell,m}(r)}{r^{a}}\right]^2$, $\left[H_{\ell,m}^{(s_3)}(\theta)\right]^2$ and $\left[G_{m}^{(s_1,s_2)}(\phi)\right]^{2}$ respectively. Since, Tsallis entropy is a moment generating function, it is defined by entropic moment; it is known analytically for $\widetilde{\a}\in\mathbb{N}-\{1\}$. We have calculated the numerical values of the Tsallis entropy for $\widetilde{\a}=1.75$ and the representative results on Tsallis entropies, $\mathcal{T}^{(\widetilde{\a})}_{n,\ell,m}(s_1,s_2,s_3)$ are graphically shown in Fig.~\ref{fig11.tsallis}, for ScH, TiH, VH and CrH molecules. As in the previous figure, here again we see some similarity between the two plots in left and right panels. The left and right panels show a shallow maximum followed by a sharp minimum and after that the entropy continually increases to reach a limiting value. In the middle panel, a sharp minimum is seen. 
As we know there exists a relation between R\'enyi and Tsallis entropies as: $\mathcal{T}^{(\widetilde{\a})}_{n,\ell,m}(s_1,s_2,s_3)=\frac{1}{\a-1}\left(1-e^{(1-\a)\mathcal{R}^{(\widetilde{\a})}_{n,\ell,m}(s_1,s_2,s_3)}\right)$. If $\mu_x=\mu_y=\mu_z=0$,  then for Tsallis entropy, we refer the reader to \cite{dehesa.ijqc2011,tsallis2,tsallis3,tsallis4}.
	
Similarly, to realize the effect of $\mu_x, \mu_y, \mu_z$ on Tsallis entropy of $H_{\ell,m}^{(s_3)}(\phi)$ and $G_m^{(s_1,s_2)}(\phi)$ we have plotted $\mathcal{T}^{(\theta,0.75)}_{\ell,m}$ and $\mathcal{T}^{(\phi,2.5)}_{m}$ with respect to the angular quantum number $\ell$ in (A) and azimuthal number $m$ in (B) for $\mu_x=-0.45$, $\mu_y=-0.4$, $\mu_z=-0.3$. The Tsallis entropy oscillates due to reflection but without reflection it is monotone and its nature depend on $\widetilde{\a}$.  

\section{Conclusions}\label{sec7.con}
In this article, we have considered Schr\"odinger-Dunkl equation in presence of Deng-Fan potential in three-dimensional spherical coordinates. The resultant radial equation is solved by the NU method using a simple novel approximation of the centrifugal term. This has been proposed by us recently and shown to be quite useful as well as effective for molecular potentials. The ro-vibrational energies and eigenfunctions are obtained in presence of Dunkl parameters, for arbitrary quantum numbers, in terms of Jacobi polynomials and hyper-geometric functions. To find analytical value of $S$, we have employed the factorization method in the Dunkl-Shro\"dinger framework. These are compared and contrasted with the non-Dunkl situation without having the reflection operators. Then $S$, expectation, Heisenberg uncertainty, entropic moment, disequilibrium, R\'enyi entropy and Tsallis entropy are obtained in analytic form. In all cases the marginal radial and angular quantities, as well as the total measures are offered. Representative illustrative results for ro-vibrational energies, Shannon entropic and moment generating functions are presented for four diatomic molecules (ScH, TiH, VH, CrH), in both ground and excited states. Numerically calculated values show negligible deviation from the semi-analytical results. For all the calculations in this work, the absolute per cent deviation remains well within 0.0001\%. To the best of our knowledge, this is the first attempt to deal with the Dunkl-Schr\"odinger equation for the important case of molecular potentials. In Schr\"odinger-Dunkl framework, all information theoretic measures are obtained with respect to the weighted Lebesgue measure. In the vanishing limit of $\mu_x, \mu_y, \mu_z$ and $\widehat{R}_x, \widehat{R}_y, \widehat{R}_z$, the information measures compare with the classical orthogonal polynomials  \cite{dehesa.fisher.sannon2005,dehesa.spherical2007,dehesa.jacobi2010,dehesa.ijqc2011,dehesa.laguerre2011,dehesa.hypergeometric2013}. The operators $\widehat{R}_x, \widehat{R}_y, \widehat{R}_z$, have direct effect on Dunkl angular momentum operator as well as on angular wave functions $H$ and $G$. The corresponding effect on information theoretic measures are seen to be oscillating, while without $\widehat{R}_x, \widehat{R}_y, \widehat{R}_z$, they appear monotone.

\section*{Acknowledgement}
AH thanks CSIR-UGC, New Delhi for JRF (09/0921(16264)/2023-EMR-I). DN thanks DST-SERB for TARE scheme (TAR/2021/000142) for financial support. AKR acknowledges partial financial support from SERB, New Delhi (sanction no. CRG/2023/004463).  

\section*{Conflict of Interest}
The authors declare that they have no known competing financial interests or personal relationships that could have
appeared to influence the work reported in this paper.
\section*{Author Contributions}
\noindent Akash Halder: Conceptualization, Investigation, Methodology, Writing – original draft, Writing – review \& editing. \\
Amlan K. Roy: Conceptualization, Investigation, Methodology, Writing – original draft, Writing – review \& editing.\\
Debraj Nath: Conceptualization, Investigation, Methodology, Writing – original draft, Writing – review \& editing.

\section*{DATA AVAILABILITY}
The data that support the findings of this study are available within the article.
\section*{ORCID}
\noindent Akash Halder: https://orcid.org/0009-0000-3548-9836\\
\noindent Amlan K. Roy: https://orcid.org/0000-0001-5555-8915\\
\noindent Debraj Nath: https://orcid.org/0000-0001-9937-7032 

\end{document}